\documentclass[iop,twocolumn]{aastex631}
\usepackage{amsmath}
\DeclareMathOperator{\sech}{sech}

\usepackage{gensymb}
\usepackage{hyperref}
\usepackage{graphicx,latexsym}

\newcommand{\abs}[1]{\left\lvert#1\right\rvert}
\newcommand\xone{x_1}
\newcommand\xtwo{x_2}
\newcommand\xthree{x_3}
\newcommand\cs{c_s}

\newcommand\Msun{\; {\rm M}_{\odot}}

\newcommand\kms{\; {\rm km}\;{\rm s}^{-1}}
\newcommand\pc{\;{\rm pc}}

\newcommand\Potunit{\;{\rm km}^2\;{\rm s}^{-2}}
\newcommand\Forunit{\;{\rm km}^2\;{\rm s}^{-2}\kpc^{-1}}

\newcommand\kpc{\;{\rm kpc}}

\newcommand\freq{\kms\kpc^{-1}}
\newcommand\slowrate{\kms\kpc^{-1}\Gyr^{-1}}

\newcommand\yr{\; {\rm yr}}
\newcommand\Myr{\;{\rm Myr}}
\newcommand\Gyr{\;{\rm Gyr}}

\newcommand\Surf{\Msun\;{\rm pc^{-2}}}

\newcommand\dunits{\Msun\;{\rm pc^{-3}}}

\newcommand\RCR{R_{\rm CR}}

\newcommand\Omb{\Omega_{b}}

\newcommand\C{{\mathcal C}}

\newcommand\inflowunit{\Msun\yr^{-1}}

\newcommand\simgt{\lower.5ex\hbox{$\; \buildrel > \over \sim \;$}}
\newcommand\simlt{\lower.5ex\hbox{$\; \buildrel < \over \sim \;$}}
\newcommand{\RNum}[1]{\uppercase\expandafter{\romannumeral #1\relax}}
\newcommand\HI{\rm H\:{{\RNum{1}}}}

\def\lvplot{($l,v$) diagram}
\def\lvplots{($l,v$) diagrams}

\def\spose#1{\hbox to 0pt{#1\hss}}
\def\dt{\spose{\raise 1.0ex\hbox{\hskip2pt$\mathchar"201$}}}

\shortauthors{Li et al.}

\begin{document}

\title{Gas Dynamics in the Galaxy: Total Mass Distribution and the Bar Pattern Speed} %% title

\author[0000-0002-0627-8009]{Zhi Li}
%\email{zli0804@sjtu.edu.cn}
\affiliation{Tsung-Dao Lee Institute, Shanghai Jiao Tong University, Shanghai 200240, P.R. China; Email: zli0804@sjtu.edu.cn}
\affiliation{Department of Astronomy, School of Physics and Astronomy, Shanghai Jiao Tong University, 800 Dongchuan Road, Shanghai 200240, P.R. China; Email: jtshen@sjtu.edu.cn}

\author[0000-0001-5604-1643]{Juntai Shen}
\correspondingauthor{Juntai Shen}
\email{jtshen@sjtu.edu.cn}
\affiliation{Department of Astronomy, School of Physics and Astronomy, Shanghai Jiao Tong University, 800 Dongchuan Road, Shanghai 200240, P.R. China; Email: jtshen@sjtu.edu.cn}
\affiliation{Key Laboratory for Particle Astrophysics and Cosmology (MOE) / Shanghai Key Laboratory for Particle Physics and Cosmology, Shanghai 200240, P.R. China}

\author[0000-0003-3333-0033]{Ortwin Gerhard}
%\email{gerhard@mpe.mpg.de}
\affiliation{Max-Planck-Institut f{\"u}r Extraterrestrische Physik, Gie{\ss}enbachstra{\ss}e, D-85748 Garching, Germany}

\author[0000-0002-2243-178X]{Jonathan P. Clarke}
%\email{jclarke@mpe.mpg.de}
\affiliation{Max-Planck-Institut f{\"u}r Extraterrestrische Physik, Gie{\ss}enbachstra{\ss}e, D-85748 Garching, Germany}

%===============================================================================

\begin{abstract}

Gas morphology and kinematics in the Milky Way contain key information for understanding the formation and evolution of our Galaxy. We present hydrodynamical simulations based on realistic barred Milky Way potentials constrained by recent observations. Our model can reproduce most features in the observed longitude-velocity diagram, including the Central Molecular Zone, the Near and Far 3-kpc arms, the Molecular Ring, and the spiral arm tangents. It can also explain the non-circular motions of masers from the recent BeSSeL2 survey. The central gas kinematics are consistent with a mass of $6.9\times10^8\Msun$ in the Nuclear Stellar Disk. Our model predicts the formation of an elliptical gaseous ring surrounding the bar, which is composed of the 3-kpc arms, Norma arm, and the bar-spiral interfaces. This ring is similar to those ``inner'' rings in some Milky Way analogs with a boxy/peanut-shaped bulge (e.g. NGC 4565 and NGC 5746). The kinematics of gas near the solar neighbourhood are governed by the Local arm. The bar pattern speed constrained by our gas model is $37.5-40\freq$, corresponding to a corotation radius of $\RCR=6.0-6.4\kpc$. The rotation curve of our model rises gently within the central $\sim5\kpc$, significantly less steep than those predicted by some recent zoom-in cosmological simulations.

\end{abstract}


\section{Introduction}

The observed strong non-circular motions of atomic and molecular gas in the central region of the Galaxy \citep{bur_lis_78,bally_etal_87,dame_etal_01} revealed the presence of a stellar bar decades ago \citep{lis_bur_80,ger_vie_86,binney_etal_91}. Many features in the observed \lvplot, which shows the the distribution of gas emission line intensity or brightness temperature as a function of Galactic longitude $l$ and line-of-sight (LOS) velocity $v$, can be explained by the periodic orbits in a barred potential. This is because the gas streamlines away from shock regions following the periodic orbits with small deviations \citep[e.g.][]{robert_etal_79,eng_ger_97,reg_teu_04,kim_etal_12a,sorman_etal_15a,sorman_etal_15b}. Thus the \lvplot\ offers tight constraints on the galactic potential, especially in the central part. Besides the \lvplot, the Bar and Spiral Structure Legacy (BeSSeL) survey \citep[][]{brunth_etal_11} obtained the position and 3D velocity of nearly two hundreds high-mass star-forming regions (HMSFRs) with high-precision Very Long Baseline Interferometry (VLBI) data \cite[e.g.][]{reid_etal_19}. The peculiar motions of HMSFRs are found to be large ($\sim50\kms$) around the bar end, which is likely related to the bar and large-scale spiral arm dynamics. 

There have been many attempts to model the global gas features in the Milky Way (MW) \citep[e.g.][]{eng_ger_99,fux_99b,bissan_etal_03,rod_com_08,baba_etal_10,pettit_etal_14,pettit_etal_15,sorman_etal_15c,li_etal_16}, but no models have yet explain all the observed $(l,v)$ features and the motion of HMSFRs simultaneously. The reason is probably due to the complexity of the Galactic structures. The Galactic bar exhibits a boxy/peanut (b/p) geometry in the central $\sim2\kpc$ \citep{dwek_etal_95,weg_ger_13,nes_lan_16,simion_etal_17}, then it gradually transitions into a long-thin bar extending to $l \approx 27-30\degree$ \citep{hammer_etal_00, benjam_etal_05, cabrer_etal_08,wegg_etal_15}. The overall shape is similar to the buckled bars that are vertically thick in the inner region seen in $N$-body simulations \citep{com_san_81, raha_etal_91, martin_etal_06,shen_etal_10,li_shen_15}. The b/p geometry generates a weaker quadrupole in the potential compared to a pure triaxial ellipsoid with the same surface density, thus may need to be taken into account in dynamical models \citep[e.g.][]{fragko_etal_16}.

Besides the bar, the effects of the large-scale spiral arms are also important for modelling gas flows \citep[e.g.][]{bissan_etal_03,seo_kim_14,pettit_etal_14}. The nearby five spiral arms, namely the Outer, Perseus, Local, Sagittarius, and Scutum Arms have been extensively studied with different tracers in recent years \citep[see the reviews by][]{xu_etal_18,shen_zheng_20}. The observed arms have pitch angles in a range of $10\degree-20\degree$, and some of them may extend up to $\sim15\kpc$ from the Galactic Center (GC) \citep{dam_tha_11}. While the shape of the spiral pattern is
relatively well-constrained, a compelling dynamical explanation for the origin of this spiral pattern in our Galaxy is still lacking \citep[but see the discussion in ][]{sellwo_etal_19a}, and different spiral driving mechanisms may have distinct effects on stars and gas.

Stellar dynamical models developed in recent years have significantly improved our understanding on the gravitational potential of the MW \citep[see the reviews by][]{bla_ger_16}. For example, the made-to-measure (m2m) models in \citet{portail_etal_17} (hereafter P17) reproduced well the observed star counts and stellar kinematics in the bulge and bar region, and many parameters of the bar (e.g. mass, pattern speed, length, axis ratio, orientation with respect to the Sun, etc.) are relatively well-constrained. Similarly, from a sample of $\sim200$ maser sources with proper motions and parallaxes, the morphology of the Milky Way's spiral arms and the kinematics for the star-forming disk are also well-constrained \citep[][]{reid_etal_19}. In addition, rotation curve measurements outside the bar radius have achieved unprecedented precision with the help of the \textit{Gaia} data \citep[e.g.][]{eilers_etal_19}. These constraints on the Milky Way's potential already provide a valuable starting point for the investigation of the gas flow in the present paper.

Our goal in this paper is to construct a gas dynamical model that can explain the observed gas kinematics (i.e. the \lvplot\ and the HMSFRs), such that we could use it to further constrain the bar pattern speed and the overall potential of the Milky Way inside the solar circle. To combine with the stellar results, we adopt the m2m potentials in P17 as inputs. We would like to see: (1) How well the m2m models agree with gas kinematics? (2) Can we use gas kinematics to provide independent and additional constraints on the bar pattern speed, $\Omb$, as well as the mass distribution of the Galaxy? We aim to provide a better gravitational potential model for the MW by combining the stellar and gas dynamics, which would be useful for many other studies.

The paper is organized as follows: we describe our Galactic potential models in \S\ref{sec:gravipot} and constraints from observations in \S\ref{sec:constrains}. The numerical methods are discussed in \S\ref{sec:numerics}. We present the constrained gas models in \S\ref{sec:overall}, the pattern speed measurements from gas dynamics in \S\ref{sec:ombdetermin}, and the related mass distributions in different regions in \S\ref{sec:massdistri}. We discuss the implications of this work and compare our results with other studies in \S\ref{sec:discussion}, and summarize in \S\ref{sec:summary}. 

%fig1
\begin{figure*}[!t]
\includegraphics[width=1.00\textwidth]{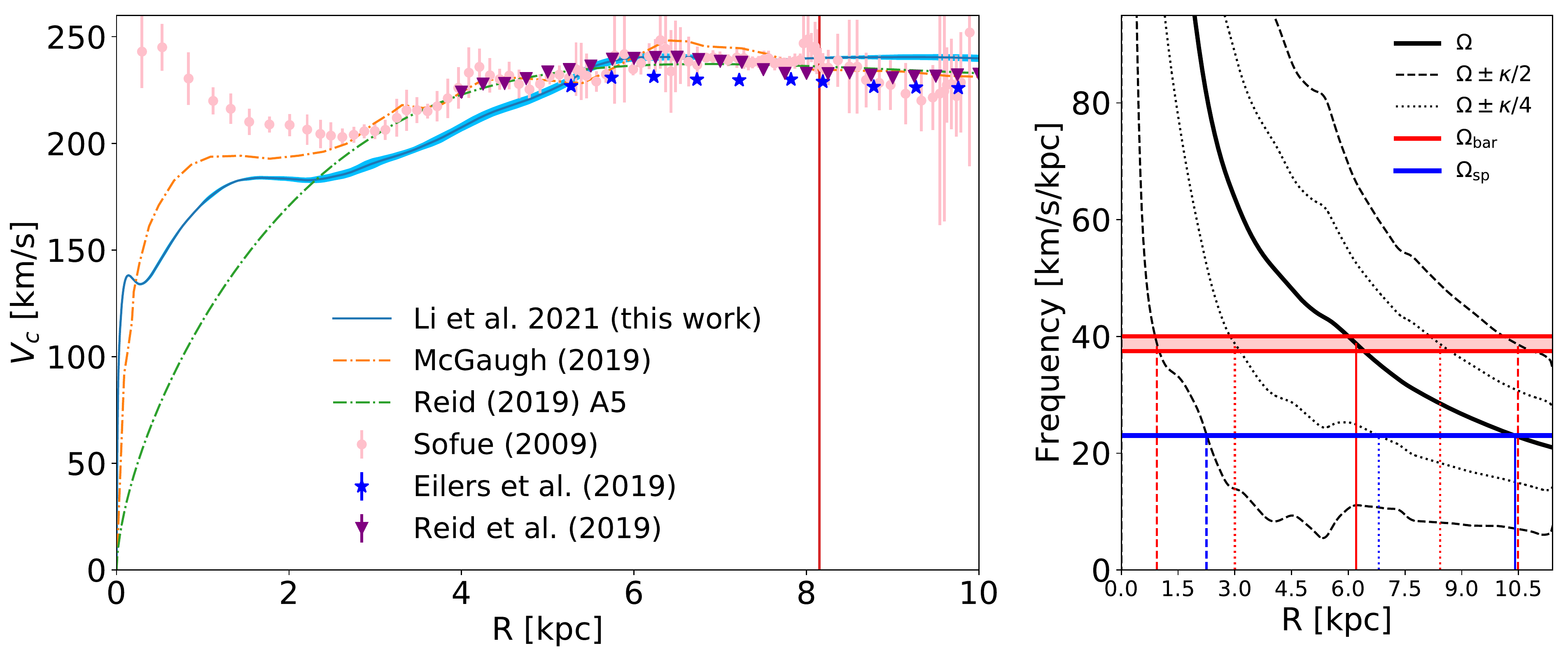}
\caption{Left panel: the circular rotation curve of the Milky Way based our fiducial gas models with $\Omb=37.5\freq$ and $\Omb=40\freq$, and the comparison with other studies. The rotation curves in the models are obtained by azimuthally averaging $\Phi_{\rm gal}$. The shaded light blue region shows the differences of the rotation curves between the model with the two bar pattern speeds. The blue line plots the middle values in the shaded region. The vertical red line indicates the adopted solar radius $R_{\odot}=8.15\kpc$ as in \citet{reid_etal_19}. Right panel: Corresponding frequency curves and resonances in our fiducial model. The solid black line $\Omega(R)$ is obtained based on the blue line in the left panel, and other black lines are calculated according to the solid black line. The horizontal two red lines indicate the two bar pattern speeds, and the vertical red lines represent the mean resonance radii of the two bar pattern speeds. The blue lines are those for the spiral pattern. 
\label{fig:rotcurve}}
\vspace{0.2cm}
\end{figure*}

% Table 1
\begin{deluxetable}{llc}
\label{table:mass}
\tablehead{
\colhead{Component} & \colhead{Mass (${\rm M}_{\odot}$)} & \colhead{Reference}}
\startdata
Nuclear Stellar Cluster & $0.61\times10^{8\ }$   & \citet{chatzo_etal_15} \\
Nuclear Stellar Disk    & $6.90\times10^{8\ }$   & \citet{sorman_etal_20b} \\
Boxy-peanut bulge       & $1.34\times10^{10}$  & \citet{portail_etal_17} \\
Long thin bar           & $0.54\times10^{10}$   & \citet{portail_etal_17} \\
Bar-spiral interface    & $0.44\times10^{8\ }$   & \S\ref{sec:bspotential} \\
4-arm spiral            & $8.39\times10^{8\ }$   & \S\ref{sec:sppotential}
\enddata
\caption{Mass of different components in our model.}
\end{deluxetable}

\section{Galactic Gravitational Potential}
\label{sec:gravipot}

We follow a similar approach to constrain the Galactic potential $\Phi_{\rm gal}$ as in our earlier work \citep{li_etal_16}. The potential $\Phi_{\rm gal}$ adopted in this work is a superposition of several components that dominate in different regions (Eq.~\ref{eq:galpot}). More specifically, we include a potential $\Phi_{\rm ns}$ of the nuclear structures, a potential $\Phi_{\rm sp}$ of two pairs of 2-arm spirals in the outer disk, a potential $\Phi_{\rm bs}$ of the bar-spiral transition interface, and a potential $\Phi_{\rm pl}$ to shift the rotation curve in a radial range of $3-7.5\kpc$. All of these components are superimposed on a basis potential $\Phi_{\rm m2m}$ from the P17 m2m models.
\begin{equation}
\label{eq:galpot}
\Phi_{\rm gal} = \Phi_{\rm m2m}+\Phi_{\rm ns}+\Phi_{\rm sp}+\Phi_{\rm bs}+\Phi_{\rm pl} 
\end{equation}
We then evolve a gas disk under such a potential and use the gas properties in the MW as observational constraints. We adjust the parameters of the above components until the combined potential $\Phi_{\rm gal}$ can generate a gas disk that reproduces most observed features. The mass of the main components in our potentials are summarized in Table.~\ref{table:mass}.

In Fig.~\ref{fig:rotcurve} we show the rotation curves and the corresponding resonances based on two $\Phi_{\rm gal}$ used in our models that can best reproduce the observations. One model has a bar pattern speed of $\Omb=37.5\freq$, and the other has $\Omb=40\freq$. We define these two models as the fiducial models in this work. The shaded regions indicate the differences of the rotation curves between these two, and the blue line is the averaged rotation curve. From the observational side, the pink dots are from \citet{sofue_etal_09} based on terminal velocities of ISM. Note the high velocity peak at $R<2\kpc$ indicated by the pink dots may not reflect the real mass distribution due to the presence of large non-circular motions of gas in this region \citep{binney_etal_91,chemin_etal_15}. The blue stars are from \citet{eilers_etal_19} based on Jeans modelling of red giants, and the orange dot-dash line is from \citet{mcgaug_19} who further included the spiral arms in the Jeans modelling to refine the rotation curve. The purple triangles and the green dot-dash lines are from \citet{reid_etal_19} based on the kinematics of HMSFRs and young stars. The lines and points seem to relatively agree with each other at $R\gtrsim5\kpc$, but there are still uncertainties inside this radius where the bar dominates the potential. In the right panel we show the corresponding frequency curves and the resonances of the fiducial models. For the model with $\Omb=37.5\freq$, the corotation radius (CR) is $6.4\kpc$, the inner Lindblad resonance (ILR) is $1.0\kpc$, the outer Lindblad resonance (OLR) is $10.8\kpc$, the inner 4:1 resonance is $3.1\kpc$, and the outer 4:1 resonance is $8.7\kpc$. For $\Omb=40\freq$ these values are $6.0$, $0.9$, $10.1$, $2.9$, and $8.2\kpc$, respectively. For the spiral arms with the pattern speed of $\Omega_{\rm sp}=23\freq$, CR is $10.4\kpc$, ILR is $2.2\kpc$, and inner 4:1 resonance is $6.8\kpc$. Note in our model the bar OLR and the spiral CR are quite close to each other. 

\begin{figure*}[!t]
\includegraphics[width=1.00\textwidth]{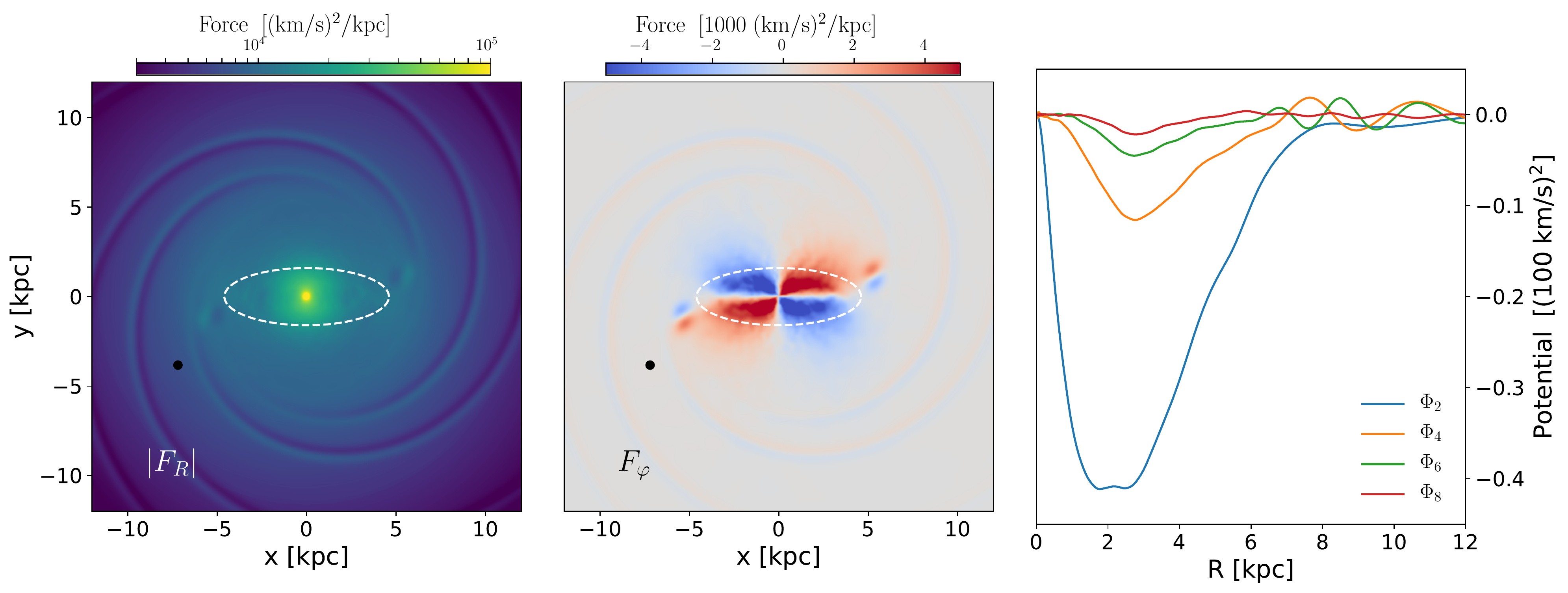}
\caption{Non-axisymmetric properties of the potential in the fiducial model with $\Omb=37.5\freq$. Left panel: the map of the radial force $F_{\rm R}$ at the midplane ($z=0$). The white dashed ellipse roughly outlines the size and shape of the bar. The black dot indicates the adopted location of the Sun. Middle panel: the map of the tangential force $F_{\rm \varphi}$ at the midplane ($z=0$) generated by the bar and spirals. The kinks around the bar ends show the bar-spiral interface described by Eq.~\ref{eq:bs}. Right panel: the quadrupole $\Phi_2$, octupole $\Phi_4$, and higher order multipoles ($\Phi_6$ and $\Phi_8$) of the potential as a function of radius.
\label{fig:frfphi}}
\vspace{0.2cm}
\end{figure*}

Fig.~\ref{fig:frfphi} shows the non-axisymmetric properties of the potential in the fiducial model with $\Omb=37.5\freq$. The left and middle panels illustrate the radial and tangential force ($F_{\rm R}$ and $F_{\rm \varphi}$) distributions at the midplane ($z=0$). The shape of the bar \& spirals, as well as their dominated regions is clearly seen in the plots. The right panel shows the multipoles of the potential obtained from Fourier decomposition. The bar leads to significant $\Phi_2$ and $\Phi_4$ components inside $R\sim5\kpc$, while the spiral contributes to the wiggles of $\Phi_4$ and $\Phi_6$ at larger radius. Multipoles with higher order than $\Phi_8$ are not important in our potentials.

In the following subsections, we explain in detail how we model the different potential components. Note the values of the parameters in \S\ref{sec:nsdnsc}-\S\ref{sec:rcpotential} are from the fiducial models that are discussed in the main part of this paper.

%fig2

\subsection{Bulge-bar-disk potential from made-to-measure modelling}
\label{sec:m2mpot}

The basis potentials used in this work are from the best-fit m2m models in P17. The m2m models were constructed from a set of $N$-body barred disks that were adiabatically adapted to match the following observed quantities. The density profiles of the m2m models are constrained by the red clump giants (RCGs) in the bulge and bar region from VVV, 2MASS, and UKIDSS; the kinematics are constrained with data from BRAVA, OGLE, and ARGOS surveys. For detailed information about the m2m models we refer the readers to P17. 

We consider four gravitational potentials from the m2m models in P17, all adjusted to the bulge/bar data assuming the same mass-to-clump ratio and nuclear stellar disk (NSD) mass, but with different bar pattern speeds of $\Omb=35.0$, $37.5$, $40.0$, and $42.5\freq$, placing bar corotation radius at $\RCR=6.72$, $6.25$, $5.83$, and $5.42\kpc$ from the GC, respectively. The mass distribution of these four models are also slightly different, resulting in different rotation curves. We average $\Phi_{\rm m2m}$ in the m2m models over a short time period to remove local fluctuations. We further impose an up-down symmetry of the potential with respect to the Galactic mid-plane ($z=0$).

\subsection{Nuclear Components}
\label{sec:nsdnsc}

The nuclear component used in P17 is an elongated exponential disc following the bar orientation with an axis ratio of 2:1. However, our gas model prefers an axisymmetric central mass distribution as the gas is observed to have nearly circular motions at $R\sim100\pc$ \citep[e.g.][]{hensha_etal_16}. We therefore replace this elongated disk with the fiducial model (model 3) in \citet{sorman_etal_20b} who used Jeans modelling to constrain the central mass distribution. This central mass model is axisymmetric and is composed of two components: the nuclear stellar cluster (NSC) and the NSD. The NSC is a compact stellar nucleus that dominates the potential of MW at $1\pc \la R \la30\pc$, while the NSD is a flattened stellar structure that governs the potential at $30\pc \la R \la300\pc$ \citep[e.g.][]{launha_etal_02,chatzo_etal_15,galleg_etal_20,sorman_etal_20b}. Their density profiles are described by the following equations:
\begin{equation}
\label{eq:nsdden}
\rho_{\rm NSC}(R,z)=\frac{(3-\gamma)M_{\rm NSC}}{4{\pi}q}\frac{a_0}{a^\gamma(a+a_0)^{4-\gamma}},
\end{equation}
\begin{equation}
\label{eq:nscden}
\rho_{\rm NSD}(R,z)=\rho_1\exp \left[-\left(\frac{a}{R_1}\right)^{n_1}\right] + \rho_2\exp \left[-\left(\frac{a}{R_2}\right)^{n_2}\right].
\end{equation}

The central density used is the sum of these two components with an additional scaling factor $\alpha$:
\begin{equation}
\label{eq:centralden}
\rho_{\rm c}(R,z)=\rho_{\rm NSC} + {\alpha}\rho_{\rm NSD}.
\end{equation}

In the above equations, $a$ is the defined as $a(R,z)\equiv\sqrt{R^2+{z^2}/{q^2}}$. The NSC density profile is from Eq.\;17 of \citet{chatzo_etal_15}, with parameters $\gamma=0.71$, $q=0.73$, $a_0=5.9\pc$, and $M_{\rm NSC}=6.1\times10^7\Msun$. The NSD density profile is obtained by deprojecting Model 2 of \citet{galleg_etal_20}, with parameters $q=0.37$, $n_1=0.72$, $n_2=0.79$, $R_1=5.06\pc$, $R_2=24.6\pc$, $\rho_1/\rho_2=1.311$, $\rho_2=1700\dunits$, and the scaling factor $\alpha=0.9$ \citep[see also Eqs. 27-30 in][]{sorman_etal_20b}. This configuration gives a total central (NSC+NSD) mass of $7.5\times10^8\Msun$, with an enclosed mass at $R=100\pc$ of $4.5\times10^8\Msun$. The corresponding gravitational potential $\Phi_{\rm ns}$ is numerically calculated based on Eq.~\ref{eq:centralden}.

\subsection{Spiral Arms}
\label{sec:sppotential}

The spiral arms are modeled with the same equations used in \citet{li_etal_16}, which are modified from those in \citet{junque_etal_13}. These equations
describe the potential of a four-armed spiral pattern following \citep{geo_geo_76,russei_03,hou_han_14,reid_etal_14,reid_etal_19}:
\begin{equation}
\label{eq:spiral}
\Phi_{\rm sp}(R,\varphi,z) =
\begin{cases}
-\zeta_{\rm sp}R\exp\{ - R/\epsilon_{\rm sp} - \abs{{k}z} \\
-\frac{R^2}{\sigma_w^2}[1 - \cos(m\varphi-f_{\rm m})]\}  & R \geq R_{\rm sp} \\
-\zeta_{\rm sp}R\exp\{ - R/\epsilon_{\rm sp} - \abs{{k}z} \\
-\frac{R^2}{\sigma_w^2}[1 - \cos(m\varphi-f_{\rm m})]\} \\
\times\exp[-(R-R_{\rm sp})^2/2\sigma_{\rm sp}^2]  & R  <   R_{\rm sp},
\end{cases}
\end{equation}
with the shape function $f_{m}$ and wavenumber ${k}$: 
\begin{equation}\label{eq:shapefunc} 
%f_{m}(R) = \frac{m}{{\tan}\;i}\ln{(R/R_{\rm n})}+\gamma,
f_{m}(R) = {m}\left(\frac{\ln{(R/R_{\rm n})}}{{\tan}\;i}+\gamma\right),
\end{equation}
and
\begin{equation}\label{eq:wavenumber} 
{k} = m/(R\tan{i}).
\end{equation}

The parameters are set to be: the amplitude $\zeta_{\rm sp}=800.0\Forunit$, the half-width of the spiral arms $\sigma_w=2.35\kpc$, the scale length of the spiral $\epsilon_{\rm sp}=3.8\kpc$, the pitch angle $i=12.5^\circ$, and the normalizing radius $R_{\rm n}=8.0\kpc$. These values are mostly taken from \citet{junque_etal_13}. We included two pairs of $m=2$ spiral, which are separated by different phase angles $\gamma_1=139.5^\circ$ and $\gamma_2=69.75^\circ$. $m$ defines the number of the spiral arms. Note that two pairs of $m=2$ spiral is not the same as a $m=4$ pattern, which results in a non-negligible $\Phi_6$ components (right panel of Fig.~\ref{fig:frfphi}). We restrict the spiral potential to be important only beyond the solar radius by tapering it off with a Gaussian form inside $R_{\rm sp}=9\kpc$ with $\sigma_{\rm sp}=1.5\kpc$, similar to the approach used in previous studies \citep[e.g.][]{kim_ost_06}. These values result a maximum potential perturbation at solar radius of $\sim630\Potunit$, slightly higher than $\sim550\Potunit$ constrained by \citet{eilers_etal_20}. The spiral potential becomes negligible inside $R_{\rm sp}-2\sigma_{\rm sp}=6\kpc$, which is close to the $\RCR$ of the bar in this study (see the left two panels of Fig.~\ref{fig:frfphi}). The total mass of the spiral is $8.39\times10^8\Msun$ by solving Poisson equation for $\Phi_{\rm sp}$, but this may be a lower limit as it is concentrated towards the plane. We rotate the spiral potential with a fixed pattern speed of $\Omega_{\rm sp}=23.0\freq$ as constrained by \citet{junque_etal_15} using open clusters and tested further in \citet{li_etal_16}. 

\subsection{Bar-spiral Interface}
\label{sec:bspotential}

It is well accepted that central bar rotates faster than outer spiral arms as can be seen in many simulations \citep[e.g.][]{quille_etal_11,minche_etal_12,hilmi_etal_20}, although the effect of the bar-spiral interaction on stellar and gas kinematics is still not clear. The bar parameters may vary when interacting with the adjacent spiral arms, but this is simply the beat pattern fluctuation between the bar and spirals. As the bar in the m2m model of P17 is in an approximately steady state, we include a perturbation near the bar ends to mimic the bar-spiral interface:

\begin{equation}
\begin{aligned}
\label{eq:bs}
\Phi_{\rm bs}(R,\varphi,z) = {} &
-\zeta_{\rm bs}R\exp\{-R/\epsilon_{\rm bs}-\abs{kz}\\
&-\frac{R^2}{\sigma_w^2}[1-\cos(m\varphi-f_{\rm m})]\}\\
&\times\exp{[-(R-R_{\rm bs})^2/2\sigma_{r}^2]}\\
&\times\exp{[-(\varphi-\varphi_{\rm bs})^2/2\sigma_{\varphi}^2]}.    
\end{aligned}
\end{equation}

This is a tapered spiral arm potential in the $R$ and $\varphi$ directions, similar as in \citet{li_etal_16}. $f_{\rm m}(R)$ and $k$ are defined by Eqs.~\ref{eq:shapefunc} and \ref{eq:wavenumber}. We adopt $m=2$, $R_{\rm n}=8\kpc$, $\gamma=36^\circ$ and $i=42^\circ$, resulting a shape that smoothly transitions from spirals into the bar ends. Other parameters are $\zeta_{\rm bs}=1300.0\Forunit$, $\epsilon_{\rm bs}=3.8\kpc$, and $\sigma_w=1.5\kpc$. The two Gaussian functions have $R_{\rm bs}=5.5\kpc$, $\varphi_{\rm bs}=12\degree$ and $192\degree$ (correspond to the two ends of the bar), $\sigma_{r}=0.5\kpc$, and $\sigma_{\varphi}=5\degree$. The above parameters are chosen to mimic a tail-like potential around the bar ends that connect the bar with the outer spirals (see Fig.~\ref{fig:frfphi}). By using such a pre-selected functional form with arbitrarily selected parameters, we would like to first understand how gas responses to a ``tailed'' barred potential, rather than aiming to model a realistic bar-spiral coupling system. The total mass of this component is $4.43\times10^7\Msun$ by solving Poisson equation for $\Phi_{\rm bs}$.

In the current paper we assume this perturbation co-rotates with the bar for simplicity. We will see in \S\ref{sec:barend} that the bar-spiral interface is important to reproduce the observed velocities of HMSFRs around the bar end.

\subsection{Modification of Rotation Curve}
\label{sec:rcpotential}

Recent studies from both gas \citep[e.g.][]{reid_etal_19} and stars \citep[e.g.][]{eilers_etal_19,mcgaug_19} have shown that the circular rotation velocity of the Milky Way reaches $v\sim230\kms$ at $R\sim4-5\kpc$ then becomes relatively flat beyond this radius. The m2m model from P17 has a slightly lower rotation velocity around this region. We therefore add an extra radial force to shift the rotation curve as follows:

\begin{equation}
\label{eq:plateau}
F_{\rm R,pl}(R,\varphi) = 
\begin{cases}
-\frac{{\Delta}v}{R}\frac{R-R_i}{R_m-R_i}\\
\times({\Delta}v\frac{R-R_i}{R_m-R_i}+2v_{\rm pl}), &  R_i(\varphi) \leq R < R_m \\
-\frac{{\Delta}v}{R}\frac{R-R_f}{R_m-R_f}\\
\times({\Delta}v\frac{R-R_f}{R_m-R_f}+2v_{\rm pl}), &  R_m \leq R < R_f \\
0,  & \text{otherwise}
\end{cases}
\end{equation}
with $R_i(\varphi)$ defined as:
\begin{equation}
R_i(\varphi) = \frac{3\kpc}{\sqrt{[1-(e\cos{\varphi})^2]}}. 
\end{equation}

Eq.~\ref{eq:plateau} corresponds to adding a linearly rising segment in the rotation curve between $R_i$ and $R_m$, and a linearly declining segment between $R_m$ and $R_f$ with the same amount of velocity shift ${\Delta}v$. The total mass of the Galaxy is therefore unaffected, since this modification only rearranges the mass distribution between $R_i$ and $R_f$. We adopt ${\Delta}v=15\kms$, $v_{\rm pl}=190\kms$, $R_m=5.5\kpc$, and $R_f=7.5\kpc$. $R_i$ is set to be an ellipse with $e=0.6$ to avoid the central bar region. The semi-major axis of the ellipse is $3.75\kpc$ and the semi-minor axis is $3.0\kpc$. The corresponding gravitational potential $\Phi_{\rm pl}$ is numerically computed based on $F_{\rm R,pl}$.

\subsection{Parameter space}

Although there are quite a lot of parameters listed in \S\ref{sec:gravipot}, many of them are fixed in the current study as these are relatively well constrained by observations and other studies (e.g. the shape of the NSD and the large-scale spirals). We run about 60 models to see how gas evolves in different potentials for different bar pattern speed, mainly focus on investigating $\Omb$ in the range of $35.0-42.5\freq$ and the velocity shift $\Delta v$ of Eq.~\ref{eq:plateau} in the range of $0-30\kms$. We then define the models that can best reproduce observations in this 2D parameter space as our fiducial models. We also test different parameters of the bar-spiral interface described by Eq.~\ref{eq:bs} in the fiducial models. The parameters varied are $\zeta_{\rm bs}$ in the range of $0-2000\Forunit$, $R_{\rm bs}$ in the range of $4.0-6.0\kpc$, and $i$ in the range of $30-50\degree$. More details of different models can be found in \S\ref{sec:ombdetermin} and the appendix.

%fig3
\begin{figure}[!t]
\includegraphics[width=0.5\textwidth]{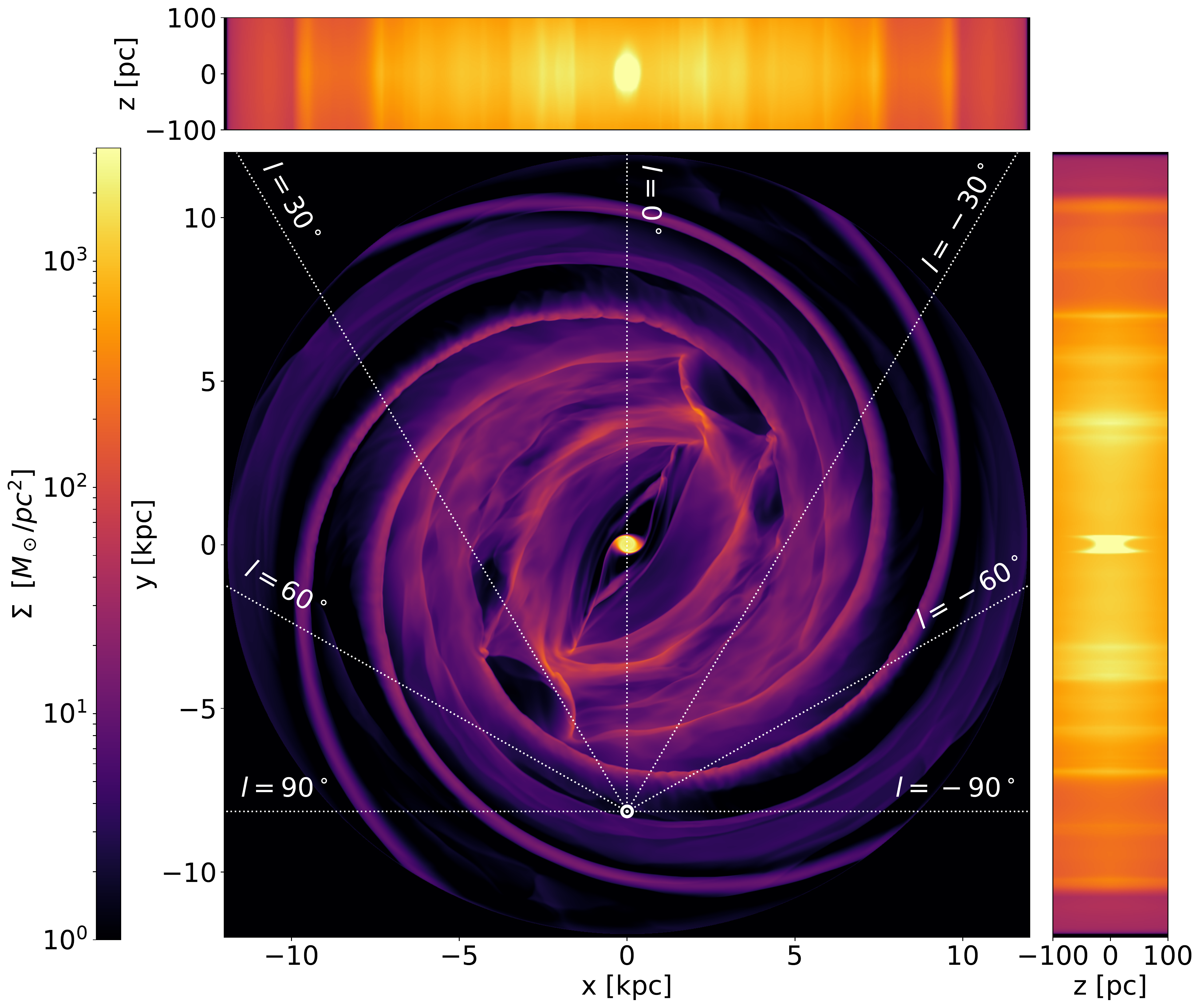}
\caption{Integrated gas surface density along three directions of our fiducial model with $\Omb=37.5\freq$ at $t=440\Myr$. Note the scale along $z$ direction is not the same as $x$ and $y$ to highlight the fine details in the vertical direction. The Sun is located at ($0\kpc$, $-8.15\kpc$) as indicated by the solar symbol. Seven white dotted lines represent different galactic longitude directions ($0\degree$, $\pm30\degree$, $\pm60\degree$, and $\pm90\degree$).
\label{fig:overall}}
\vspace{0.2cm}
\end{figure}

%fig4
\begin{figure*}[!t]
\includegraphics[width=1.00\textwidth]{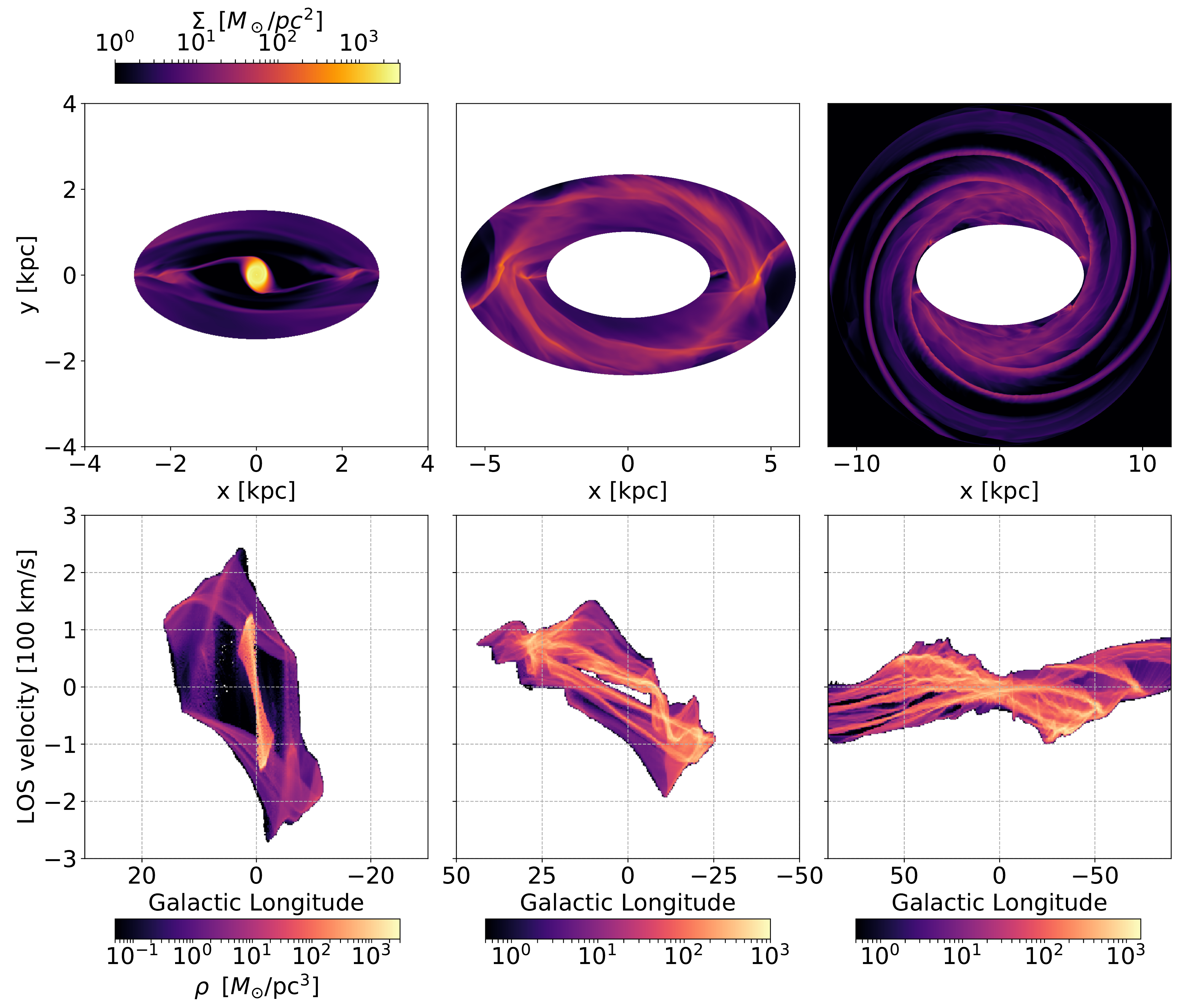}
\caption{Gas surface density plots (top tow) and the corresponding \lvplots\ (bottom row) of the fiducial model with $\Omb=37.5\freq$ in different regions. From left to right: gas flows inside an ellipse with a semi-major axis of $2.85\kpc$ and a semi-minor axis of $1.5\kpc$, the CMZ can be seen as the high-density nuclear ring/disk in the top panel and the parallelogram in the bottom panel within $\abs{l}\le2\degree$; gas flows outside the previous ellipse but inside the second ellipse with a semi-major axis of $5.83\kpc$ and a semi-minor axis of $3.5\kpc$; gas flows outside the second ellipse. 
\label{fig:surflv}}
\vspace{0.2cm}
\end{figure*}

\section{Constraints from observations}
\label{sec:constrains}

\subsection{($l,v$) diagram}
\label{sec:lvconstraints}

We compare the features in $(l,v)$ space with those identified in observed $\HI$ and CO from \citet{reid_etal_16} together with \citet{rod_com_08}, and compare terminal velocities with measurements from \citet{mcc_dic_07}, \citet{bur_lis_93}, and \citet{clemen_85}, similar as in \citet{li_etal_16}.

To obtain the \lvplot\ of the models, we assume the Sun locates at $(x,y,z)=(0\kpc,-8.15\kpc,0.025\kpc)$, with a circular velocity of the local standard of rest (LSR) to be $v_{\rm LSR} = 236\kms$. We also consider the peculiar motions of the Sun with respect to the LSR as $(U,V,W)=(10.6,10.7,7.6)\kms$. These values are taken from \citet{reid_etal_19} (see also \citealt{schonr_etal_10,bla_ger_16,gravit_etal_19}). The bar angle to the Sun-GC line is fixed to be $28\degree$ \citep{portail_etal_17}. We then calculate the Galactic longitude $l$ and line-of-sight velocity $v_{\rm LOS}$ in each cell of the simulation grid based on the solar position and the gas velocity $\vec{v}$. The results are binned onto a $(l,v)$ grid weighted by the gas density $\rho$. In the current study we do not consider radiative transfer effects, but just assume that the observed flux intensity is proportional to the gas density. The $(l,v)$ grid for the simulation has a bin size of $\Delta l = 0.6\degree$ and $\Delta v = 3\kms$.

\subsection{HMSFRs}
\label{sec:maserconstraints}

The \lvplot\ contains the information of directions and LOS velocities only. We further consider the positions and tangential motions of the HMSFRs with high-precision VLBI data \citep{reid_etal_14,reid_etal_19,veraco_etal_20} as constraints. A successful gas dynamical model should explain the \lvplot\ and the 3D motions of the observed HMSFRs simultaneously. For the current study we select the HMSFRs from \citet{reid_etal_19} that lies within $\abs{z}\le0.1\kpc$ from the Galactic plane and $R\le12\kpc$ from the GC to constrain our gas models. Furthermore, we only consider the HMSFRs that have distance uncertainties less than $0.5\kpc$. This selection results in a sample of 135 HMSFRs. As the gas in our models cannot form stars (or HMSFRs), we then interpolate our models and compare the gas properties at the same locations with the observed HMSFRs.

\section{Numerical Scheme}
\label{sec:numerics}

The simulations are performed using a modified version of {\tt Athena++} code
\footnote{\hyperlink{https://princetonuniversity.github.io/athena/}
{https://princetonuniversity.github.io/athena/}} \citep{white_etal_16,stone_etal_19,stone_etal_20}. We adopt a uniform Cartesian grid with $2048\times2048\times21$ cells covering a simulation box of $(24\times24\times0.2)\kpc$ along $(x,y,z)$ directions, respectively. The corresponding resolution is roughly $10\pc$. The typical timestep for integration is ${\Delta}t\sim1.4\times10^3\yr$. Other options adopted are piecewise linear reconstructions, the {\tt roe} Riemann solver \citep{roe_81}, and the outﬂow boundary condition. 

We essentially study the response of a thin gas disk under a realistic bar-spiral Milky Way potential described by Eq.~\ref{eq:galpot}. The initial gas disk has a density profile:

\begin{equation}\label{eq:surfden}
\rho_{\rm gas}(R,z) = \frac{\Sigma_{\rm g}}{2z_{\rm gas}}\exp{(-R/R_{\rm gas})}\sech^2{(z/z_{\rm gas})},
\end{equation}
where $\Sigma_{\rm g}=71.1\Surf$, $R_{\rm gas}=4.8\kpc$, and $z_{\rm gas}=130\pc$. The corresponding gas surface density at the solar radius ($R_{\rm \odot}=8.15\kpc$) is $\Sigma_{\rm g,0}=13\Surf$, consistent with \citet{bov_rix_13} and P17. The total mass of the gas disk is $\sim10^{10}\Msun$. 

We start with gas on circular orbits in an axisymmetrised potential, which is obtained by azimuthally averaging $\Phi_{\rm gal}$. The non-axisymmetric components are linearly ramped up during the first $100\Myr$ to avoid transients, similar to previous studies \citep[e.g.][]{kim_etal_12a,sorman_etal_15a}.

We adopt an isothermal Equation of State (EoS) and assume an \textit{effective} isothermal sound speed $\cs=10\kms$, same as in previous studies \citep[e.g.][]{fux_99b,rod_com_08,kim_etal_12a,ridley_etal_17}. This sound speed makes gas transition from the $x_1$ to $x_2$ orbits happens in the places consistent with observations \citep[][]{sorman_etal_15a}. The corresponding temperature is $15422\;{\rm K}$ assuming a mean molecular weight of 1.273 \citep[e.g.][]{glo_cla_12}. Note this effective sound speed reflects the velocity dispersion of gas clouds in the Galactic disk \citep[around $10\kms$, see][]{burton_76,walter_etal_08,tambur_etal_09} instead of a microscopic temperature.
The isothermal EoS enables us to focus more on the effects of the non-symmetric potential on gas flows by neglecting microscopic physics like cooling, star formation, and stellar/SN feedback. It also helps to explore a larger parameters space of the potential as the current simulation setup is not very computationally expensive (it takes about 3 days to evolve one model to $500\Myr$ using 256 Intel Xeon(R) Gold 6240 cores). The self-gravity of gas is neglected. We regard our isothermal simulations as a ﬁrst-order approximation to the observed cold gas in the Milky Way. We also present the models with different $\cs$ in Figs.~\ref{fig:cs5} and \ref{fig:cs15}. The caveats of this assumption are discussed in \S\ref{sec:futurework}.

%fig5
\begin{figure*}[!t]
\includegraphics[width=1.0\textwidth]{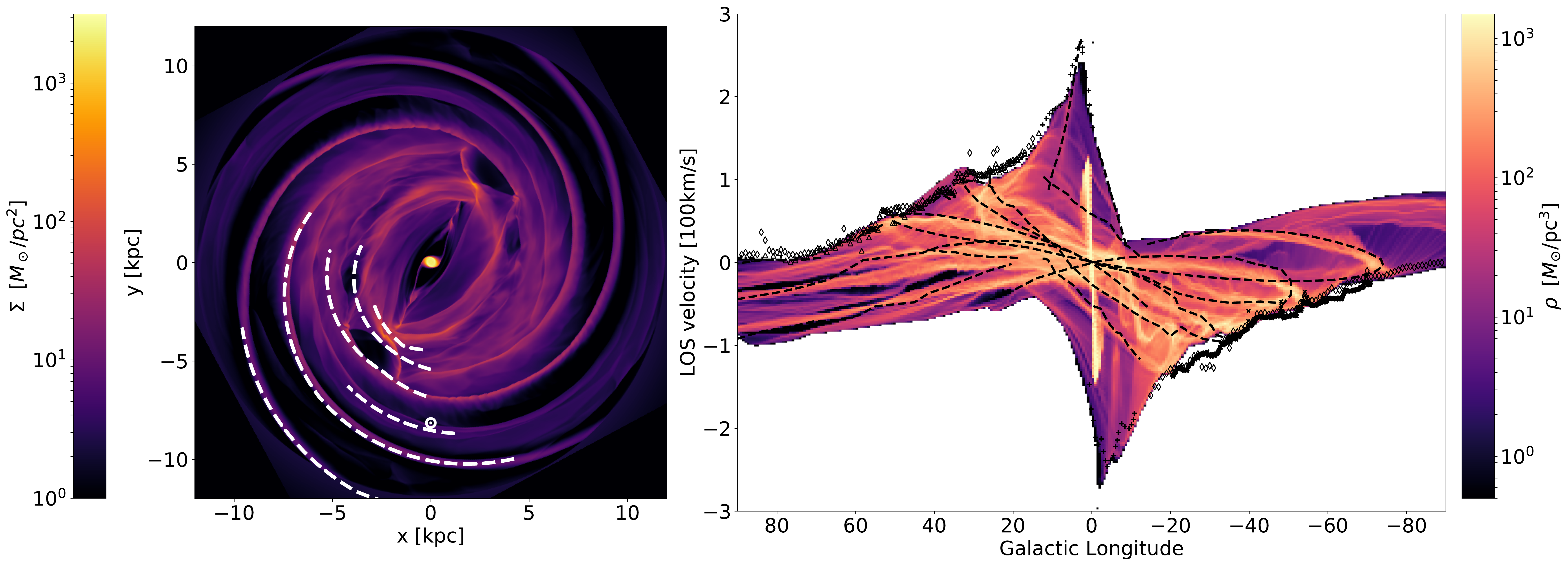}
\caption{Our fiducial gas model with $\Omb=37.5\freq$. Left panel: the gas surface density map. The solar symbol represents the Sun's location at $(0\kpc$, $-8.15\kpc$). The bar major axis has a angle of $28\degree$ with respect to the Sun-GC line. The white dashed lines show spiral arm locations from \citet{reid_etal_19}. Right panel: the \lvplot\ of the model. Dashed lines represent various features identified in \citet{reid_etal_16} and \citet{rod_com_08}. The limits of the color bar are chosen to highlight most of the features. The black crosses, diamonds, and plus signs are $\HI$ terminal velocities adopted from \citet{mcc_dic_07}, \citet{fich_etal_89}, and \citet{bur_lis_93}, respectively. The open triangles show the CO terminal velocities from \citet{clemen_85} at positive longitudes.
\label{fig:Omb37}}
\vspace{0.2cm}
\end{figure*}

%fig6 
\begin{figure*}[!t]
\includegraphics[width=1.0\textwidth]{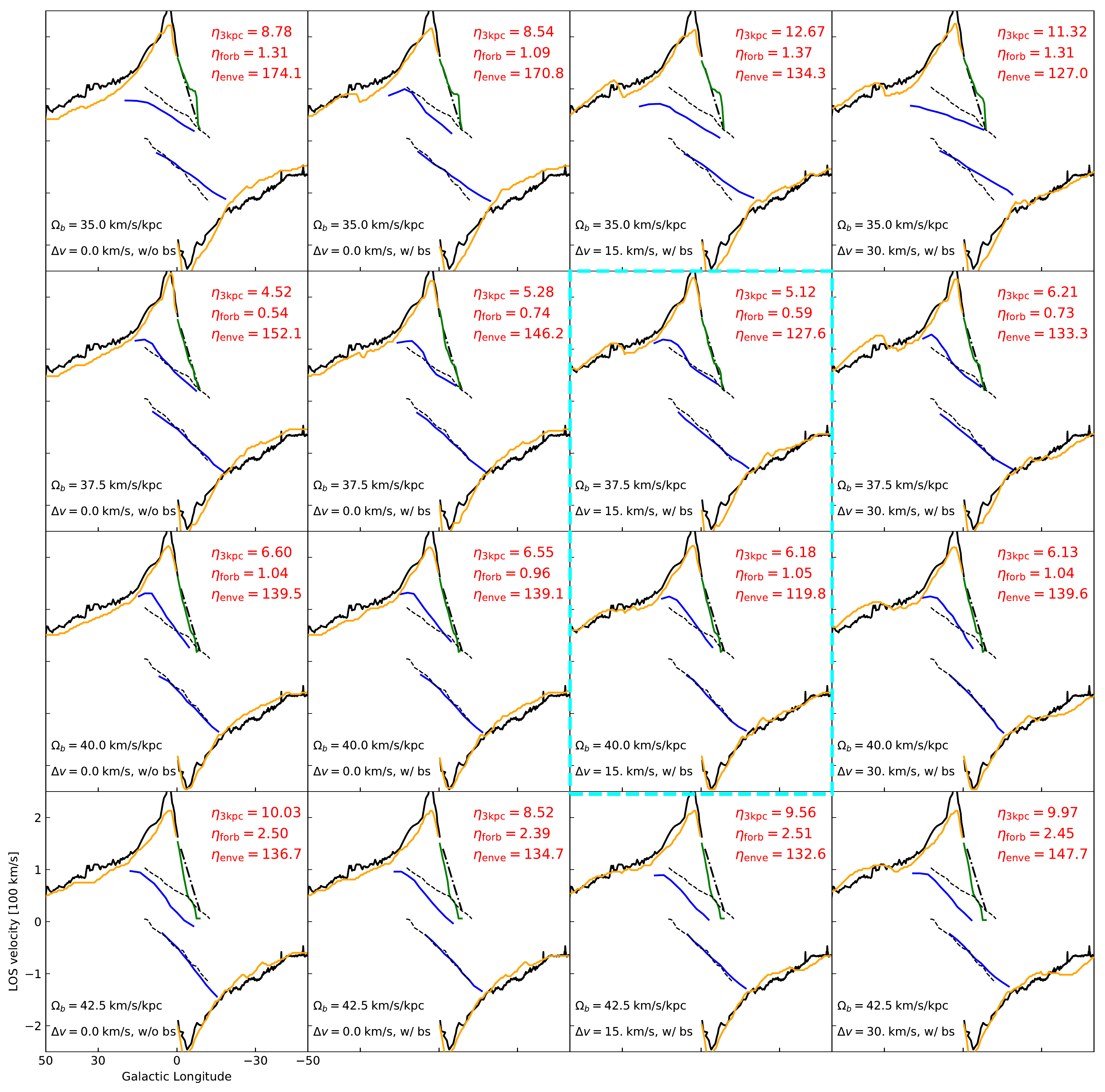}
\caption{Comparison between models. From left to right: gas model in the potential with neither the bar-spiral interface nor the rotation curve shift; gas model with the bar-spiral interface but without the rotation curve shift; gas model with the bar-spiral interface and the rotation curve shift with $\Delta v=15\kms$; gas model with the bar-spiral interface and the rotation curve shift with $\Delta v=30\kms$. From top to bottom: gas model with $\Omb=35$, $37.5$, $40$, and $42.5\freq$. In each panel, the black solid lines represent the terminal velocities from \citet{clemen_85}, \citet{fich_etal_89}, \citet{bur_lis_93}, and \citet{mcc_dic_07}; the black dashed lines represent the 3-kpc arms from \citet{reid_etal_16}; and the black dot-dash line represents the forbidden velocity from \citet{rod_com_08}. The yellow, blue, and green solid lines are the corresponding features in the models. The differences between the model and the observations for these three features are listed on the upper right of each panel. The two fiducial models are highlighted by the cyan dashed box. 
\label{fig:3kpccomp}}
\vspace{0.2cm}
\end{figure*}

\section{Overall gas morphology and kinematics}
\label{sec:overall}

Fig.~\ref{fig:overall} shows the integrated gas surface density in our fiducial model with $\Omb=37.5\freq$ and $\Delta v=15\kms$. We chose the location of the Sun (white solar symbol in the plot) based on a fixed bar angle to the Sun-GC line of $28\degree$, which is well-constrained by P17. This snapshot is taken when the spirals roughly finish one rotation period in the bar co-rotating frame, i.e. $T_{\rm p}=2\pi/\Delta\Omega\approx440\Myr$, where $\Delta\Omega=\Omb-\Omega_{\rm sp}$. As we initialized the spiral potential based on the observed spiral phase angles in \citet{reid_etal_19} (see Fig.~\ref{fig:frfphi}), both the bar and spiral arms have appropriate locations relative to the Sun around $T_{\rm p}=440\Myr$ as the initial state.

In the central $\sim200\pc$ of the plot there is a high-density nuclear gas ring/disk, which corresponds to the observed Central Molecular Zone (CMZ). Two strong shocks emerge from the central disk and roughly extend to the bar ends. This is a typical gas flow pattern in a rotating barred potential \citep[e.g.][]{athana_92b,kim_etal_12a,sorman_etal_15a,li_etal_15}. At larger radius an elliptical gaseous ring surrounding the bar ends around $l\sim\pm30\degree$. This type of gas ring around the bar dubbed an ``inner ring'', has also been commonly found in external barred galaxies \citep{vaucou_59,sandag_61,combes_96,kor_ben_19} and in cosmological hydrodynamical simulations \citep{grand_etal_17,fragko_etal_20}. A few gas spurs connect the ring to the spiral arms in the outer disk regions. The Sun is located slightly inside the Local arm, which also shows a well-defined spiral shape in the plot. These gas structures could have also been identified from the edge-on view as ridge-lines.

Fig.~\ref{fig:surflv} presents a clearer view of the gas in different regions and their corresponding kinematics in the $(l,v)$ space. Note in the top row the bar is always aligned with the $x$-axis for better illustration. We separate the gas flow pattern into three parts: the bar-driven inflows (left column), the ``inner ring'' around the bar (middle column), and the large-scale outer spirals (right column). The flow in the left column is highly non-circular, and the gas can reach a LOS velocity of $\geq200\kms$ since it flows in streams along the shocks. The typical LOS velocities in this region is around $100-200\kms$. The CMZ in our model corresponds to the parallelogram in the $(l,v)$ space within $\abs{l}\le2\degree$. The elliptical gas ``inner ring'' surrounding the bar in the middle column is composed of four spiral segments, and they form four lines in the $(l,v)$ space with typical LOS velocities around $50-100\kms$. The near and far 3-kpc arms in our model correspond to the top and bottom lines. The two gas spurs at the bar ends can be seen in the $(l,v)$ space as the kinks around $l\sim30\degree$ and $-15\degree$, respectively. The outer spiral arms in the right columns are shown by the high-density ridges in the $(l,v)$ space with low LOS velocities of $\sim10-50\kms$. A combined figure with observations included is presented by Fig.~\ref{fig:Omb37}. We explain in the following sections how we constrain the bar pattern speed, together with the galactic potential, by directly comparing these structures to observations.

\section{Bar pattern speed and the rotation curve from gas dynamics}
\label{sec:ombdetermin}

The \lvplots\ of the models with four bar pattern speeds ($\Omb=35.0$, $37.5$, $40.0$, and $42.5\freq$), three rotation curve shifts (${\Delta}v$ from Eq.~\ref{eq:plateau} with $0$, $15$, and $30\kms$), and the presence (or not) of the bar-spiral interface $\Phi_{\rm bs}$, are summarized in Fig.~\ref{fig:3kpccomp}. We only plot the terminal velocity (i.e. the envelop of the \lvplot), the 3-kpc arms, and the forbidden velocity region in order to highlight the differences between the models in the $(l,v)$ space. The 3-kpc arms are the bar-driven spirals discussed in \S\ref{sec:overall}, and the forbidden velocity regions are defined as ($l\le0\degree$, $v_{\rm LOS}\ge0\kms$) and ($l\ge0\degree$, $v_{\rm LOS}\le0\kms$) where gas on circular orbit is not expected to appear in these regions. The 3-kpc arms and forbidden velocities are closely related to the bar pattern speed \citep[see also][]{sorman_etal_15c}, while the terminal velocity is both associated with the bar/spiral and the shape/amplitude of the rotation curve.

Fig.~\ref{fig:3kpccomp} shows the 3-kpc arms (blue solid lines), the forbidden velocity region (green solid line), and the terminal velocity (yellow solid lines) in our models with parameters mentioned above. The corresponding features from observations are shown by the black lines. The model parameters are listed on the bottom left of each panel. Note the 3-kpc arms in the models are manually extracted from the \lvplot\ in order to avoid contamination from other features. We see from top to bottom that the 3-kpc arms become more tilted, and the forbidden velocity region is less pronounced compared to observations with a higher $\Omb$. We only plot the forbidden velocity on the negative longitude as the data is limited on the other side. The terminal velocity curves in the leftmost column are clearly below the observed one within $20\degree\la|l|\la40\degree$, which motivate us to shift the rotation curve in this region. This also implies that the disk mass inside the bar region may be underestimated in the original P17 potentials. The models with $\Delta v=15\kms$ seem enough to explain the observed terminal velocity, while the models with $\Delta v=30\kms$ produce higher curves beyond $|l|\sim 40\degree$. We also find that including the bar-spiral interface does not significantly affect the 3-kpc arms and the forbidden velocity, but it helps to create an obvious tangent around $l=30\degree$, which better agrees with observations.

To quantify the difference between models, we define $\eta_{\rm 3kpc}$ as the areas between the 3-kpc line segments in the models and observations, i.e. it is the space enclosed by the blue solid lines and dashed black lines in Fig.~\ref{fig:3kpccomp}. $\eta_{\rm forb}$ is defined in a similar way for the forbidden velocity line segment. For the terminal velocity curves, we use Dynamic Time Warping (DTW) distance $\eta_{\rm enve}$ between models and observations to capture the trend of tangents variations along Galactic Longitude. The calculations are done with the help of {\tt similaritymeasures} package \citep{jekel_etal_19} in {\tt python}. The results are listed on the upper right of each panel, and a smaller $\eta$ means a better match for a certain feature. We find that our gas models prefer a bar pattern speed within $\Omb=37.5-40\freq$ based on the diagnostics of $\eta_{\rm 3kpc}$ and $\eta_{\rm forb}$, as these parameters reach a local minimum in this $\Omb$ range. The terminal velocity prefers a rotation curve shift around $\Delta v=15\kms$ based on $\eta_{\rm enve}$. These selection criteria lead to the two fiducial models (highlighted by the cyan dashed box) mentioned in the previous sections. The corresponding $\RCR$ for these two models is in the range of $6.0-6.4\kpc$ as shown in Fig.~\ref{fig:rotcurve}. These values are also consistent with some of the previous gas dynamical models \citep[e.g.][]{rod_com_08,sorman_etal_15c} and the independent stellar kinematics measurements in the bulge \citep[e.g.][]{clarke_etal_19,sander_etal_19}.

Recent works by \citet{chiba_etal_20} suggested the Galactic bar may currently be in a decelerating phase with a slowing rate of $\dot{\Omb}=-4.5\pm1.4\slowrate$. The authors also concluded models with a constant $\Omb$ may give qualitatively wrong conclusions. We argue that our pattern speed measurement is not sensitive to a small bar slowing rate, as the response of gas flows to the bar potential is quite rapid. In our models the gas reaches quasi-steady state within about two bar rotation periods, i.e. $\sim300\Myr$ for $\Omb=37.5-40\freq$. Adopting $\dot{\Omb}=-4.5\slowrate$, this yields an uncertainty of $1.35\freq$, which is not significant. We have also tested a decelerating bar with $\Omega_b(t)=39.5-4\times(t/{\rm Gyr})$, and the resulting \lvplot\ is almost indistinguishable compared to Fig.~\ref{fig:Omb37}. On the other hand, \citet{hilmi_etal_20} predicted the $\Omb$ and the bar length may change by $20\%$ due to the bar-spiral interaction within one bar rotation. This timescale is shorter than that required by gas to reach quasi-steady state, as such our $\Omb$ measurements can only be regarded as a time-averaged value. However, we note the spirals connected with the bar seem quite massive in \citet{hilmi_etal_20}, which can be sufficient to affect the bar dynamics. For a comparison, the mass of the bar-spiral interface used in our model is $\sim4.43\times10^7\Msun$. This mass is enough to explain the large peculiar motion of HMSFRs around the bar end (see \S\ref{sec:barend} and Fig.~\ref{fig:bseffect}), but is only $\sim0.24\%$ of the total bar mass ($1.88\times10^{10}\Msun$ in P17), and $\sim0.82\%$ of the long-bar mass ($5.4\times10^9\Msun$ in P17, see also \citealt{wegg_etal_15}). The total mass of the 4-arm spiral plus the bar-spiral interface is $\sim8.83\times10^8\Msun$, which is $\sim4.69\%$ of the total bar mass. Such a mass contrast makes it unlikely that the adopted bar-spiral interface can dramatically affect $\Omb$. Future investigations are required to better constrain $\dot{\Omb}$ in the short and long term for the Milky Way's bar.

%fig7 
\begin{figure*}[!t]
\includegraphics[width=1.0\textwidth]{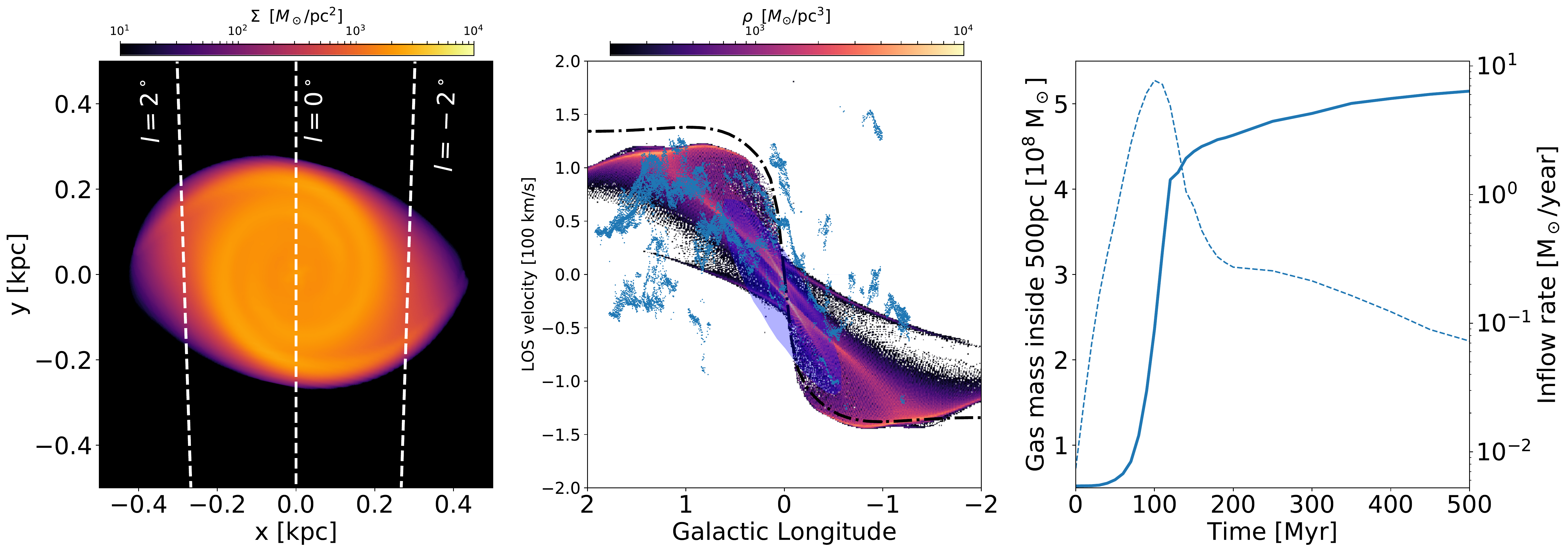}
\caption{Zoom-in view of the central region in the fiducial model with $\Omb=37.5\freq$ at t=$440\Myr$. The limits of the colorbar are adjusted compared to Fig.~\ref{fig:Omb37} to better visualize the high density region. Left panel: the gas surface density map. Middle panel: the corresponding \lvplot. We interpolate our mode with a $2\pc$ resolution within $1\kpc$ to better show the central features. The bin sizes are $\Delta l = 0.01\degree$ and $\Delta v = 1.0\kms$. The black dash-dot line indicates the circular rotation velocity adopted in the model. The blue shaded region shows the observations of ${\rm C}{\rm \RNum{2}}$ obtained from \citet{langer_etal_17}. The small blue dots are NH$_3$ observations from HOPS survey after processing with SCOUSE \citep{hensha_etal_16,longmo_etal_17}. Right panel: accumulated gas mass inside $500\pc$ as a function of simulation time (solid line), and the corresponding mass inflow rate (dashed line). 
\label{fig:cmz}}
\vspace{0.2cm}
\end{figure*}

\section{Mass distribution in different regions}
\label{sec:massdistri}

The gas morphology and kinematics provide key information for the underlying mass distribution, and could help us to better understand the large scale structures of the Galaxy. We use our fiducial model with $\Omb=37.5\freq$ as an example to discuss its gas properties and the related mass components in three different regions of the MW.

\subsection{Nuclear Stellar Disk and Central Molecular Zone}
\label{sec:cmz}

The NSD is a flattened stellar structure located in the central $R \leq 150-200\pc$ of the Galaxy, with a scale length of $\sim100-200\pc$ and a scale height of $\sim45\pc$ \citep{launha_etal_02,nishiy_etal_13,schonr_etal_15,galleg_etal_20}. The gaseous counterpart of the NSD is probably the CMZ, which is a high-density gas ring/disk with a radius of $\sim200\pc$ \citep{molina_etal_11,hensha_etal_16}. Pioneering works led by \citet{binney_etal_91} interpret the CMZ as cold gas switching from $\xone$ orbits to $\xtwo$ orbits, while for warm gas the $\xtwo$ orbits would not be occupied \citep{eng_ger_97}. The CMZ can therefore be regarded as a counterpart to circum-nuclear (or nuclear) rings in external barred galaxies \citep{comero_etal_10,li_etal_15}. The latest studies that include more physics can form structures reminiscent of the CMZ and NSD in a barred Milky Way model \citep[e.g.][]{seo_etal_19,bab_kaw_20,tress_etal_20,sorman_etal_20a}.

The gas kinematics in the galactic central region offer tight constraints on the NSD properties. A more massive and/or compact NSD may result in a steeper rise in LOS velocity compared to observations, and vice versa \citep[see][]{li_etal_20}. We show in Fig.~\ref{fig:cmz} that a central NSC+NSD mass of $7.5\times10^8\Msun$ can generate a gas disk with a similar size and kinematics as the observed CMZ. The figure shows the gas surface density and the \lvplot\ in the central $500\pc$ of our fiducial model with $\Omb=37.5\freq$. The gas ring/disk (or a pair of tightly-wounded spirals) at a radius of $\sim200\pc$ in the left panel corresponds to the observed CMZ, similar to previous simulations \citep[e.g.][]{kim_etal_11,ridley_etal_17,armill_etal_19}. The \lvplot\ of the model is shown in the middle panel. The black dash-dot line denotes the circular rotation curve of the model. The envelope of the gas \lvplot\ roughly matches the rotation curve, suggesting gas motions can be used to trace the central potential. The spikes with low velocities are formed by the gas above the midplane. The slightly up-down asymmetry of the \lvplot\ in the model is due to the $\sim10\kms$ inward solar motion mentioned in \S\ref{sec:lvconstraints}, this asymmetry is probably also noticeable in the ${\rm C}{\rm \RNum{2}}$ observation (blue shared region). The NH$_3$ data (small blue points) reach $v_{\rm LOS}=90-110\kms$ around ${l}\sim1\degree$, and the model follows a similar pattern, indicating the NSD mass profile we adopt is reasonable. The asymmetric NH$_3$ distribution with respect to $l=0\degree$ is probably due to the transient instabilities developed by the inflowed gas \citep{wad_kod_04,sorman_etal_17a}. These instabilities may also help explain the velocity drop in NH$_3$ from $\sim100\kms$ to $\sim50\kms$ in the range of $l\sim1\degree$ to $2\degree$ \citep[e.g.][]{li_etal_20}. It is also possible that the drop is created by collisions between gas streams moving at different speeds along the shocks \citep{sorman_etal_19}. Such a velocity drop is even steeper than Keplerian thus it must be highly non-circular, and it is difficult to use this drop to constrain the mass distribution. In general, we find the gas kinematics in the CMZ of our model agree with a $\sim7\times10^8\Msun$ NSD in the central $|l|=2\degree$. Interestingly, P17 obtained a dynamical central mass of $\sim2\times10^9\Msun$ based on stellar proper motions in the central region. This is about 3 times larger than our model and \citet{sorman_etal_20b}. A likely reason for this discrepancy is that the NSD mass in P17 was constrained by OGLE proper motions at higher latitudes ($\abs{b}>=2\degree$), reflecting an enclosed mass on larger scales. Further observations like \textit{JASMINE} \citep{gouda_12} and \textit{GaiaNIR} \citep{hobbs_etal_19} will provide more insight on the origin and mass of the NSD in our Galaxy. 

The right panel of Fig.~\ref{fig:cmz} shows the enclosed mass in the central $500\pc$ (solid line) and the corresponding mass inflow rate (dashed line) as a function of time. It is clear that most inflow happens in the first $100\Myr$, even less than one bar rotation period. After $\sim200\Myr$ the mass inflow rate becomes relatively small. The short inflow timescale is also revealed in star formation history (SFH) derived by \citet{noguer_etal_19}. The inflow rate at $t\ga400\Myr$ falls below $\sim0.1\inflowunit$, because gas is not replenished (or slowly by the outer spiral arm) inside the bar region of our isolated models. For a comparison, the mass inflow rate to the CMZ region based on observational estimation gives $\sim0.8-2.7\inflowunit$ \citep{sor_bar_19,hatchf_etal_21}. The large observed inflow rate may be related to the recent disk perturbation due to (massive) satellites \citep[e.g.][]{antoja_etal_18,blandh_etal_19,laport_etal_19,li_shen_20}, or possibly due to the in-falling gas clouds from the Molecular Ring reservoir perturbed by spiral arms \citep[e.g.][]{bissan_etal_03}. At the end of our simulation ($t=500\Myr$) the enclosed gas mass reaches $5.15\times10^8\Msun$. This suggests the bar is able to drive around $5\%$ of the initial gas disk ($1.0\times10^{10}\Msun$) into the center if there is little gas replenishment inside the bar region. However, the observed mass of molecular gas in the CMZ is around $\sim3-7\times10^7\Msun$ \citep{launha_etal_02,molina_etal_11}, which is about an order of magnitude lower than that in the model. This discrepancy can be understood as most dense gas would turn into stars then become part of the NSD, and a fraction of the accumulated gas would be ejected by stellar feedback, but these processes are not yet included in the current model. In addition, magnetic fields \citep{mangil_etal_19} and turbulence \citep{salas_etal_20} may also be important in modelling the detailed gas properties in the CMZ. We would like to study these effects and make a more careful comparison with the CMZoom survey \citep{hatchf_etal_20} in a future work. 

%fig8 
\begin{figure*}[!t]
\includegraphics[width=1.0\textwidth]{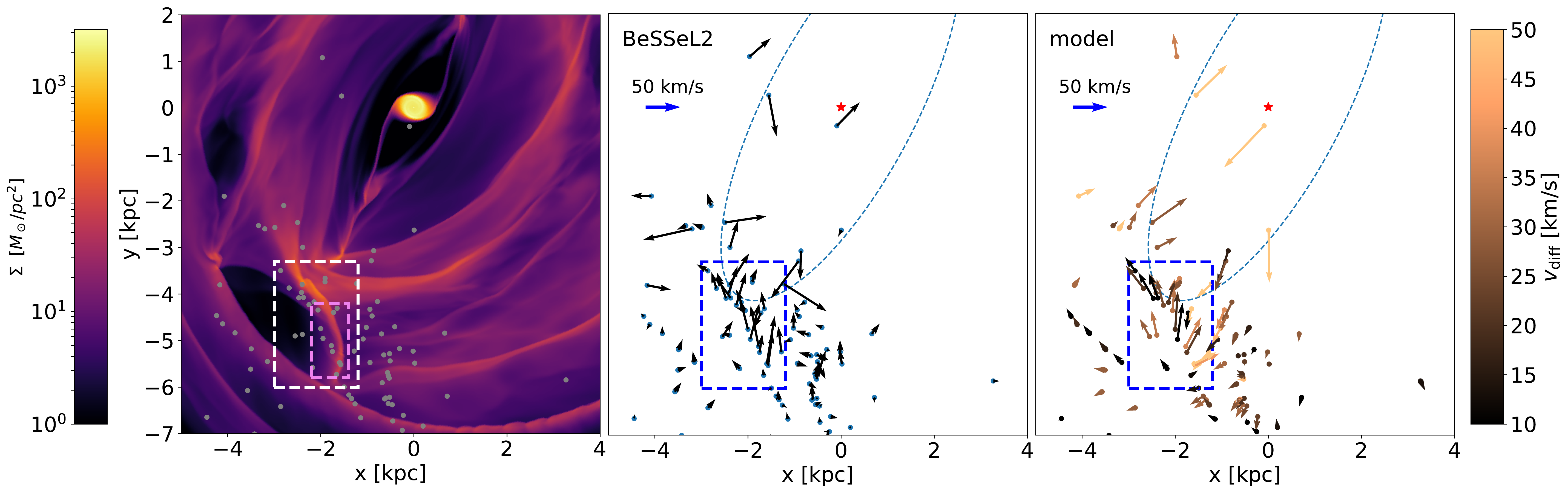}
\caption{Comparison between our fiducial model with $\Omb=37.5\freq$ and BeSSeL2 results. Left panel: the gas surface density overlaid with the locations of observed HMSFRs (grey dots) from \citet{reid_etal_19}. The white dashed box highlights the region of the observed HMSFRs with large peculiar motions. The gas spur has a mass of $1.94\times10^7\Msun$ inside the purple dashed box. Middle panel: non-circular (peculiar) motions of the HMSFRs in BeSSeL2. The blue dashed curve roughly outlines the shape of the bar. The blue dashed box shows the same region as the white one in the left. A $50\kms$ velocity vector is shown for reference at the upper left corner. The red star indicates the GC. Right panel: non-circular motions at the same locations in the model. The color of the arrow represents the velocity difference (Eq.~\ref{eq:vdiff}) between the model gas and the observed HMSFRs.)  
\label{fig:bessel}}
\vspace{0.2cm}
\end{figure*}

%fig9
\begin{figure*}[!t]
\includegraphics[width=1.0\textwidth]{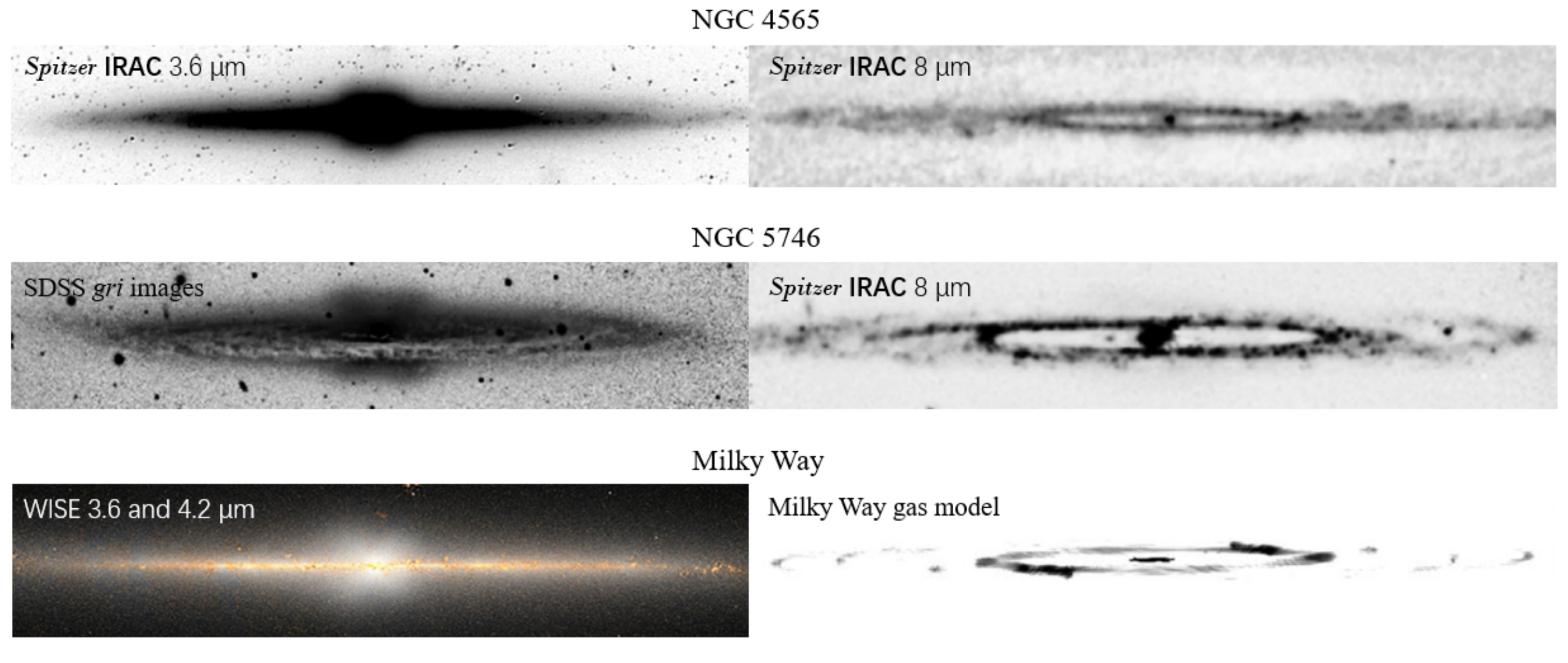}
\caption{Comparison between our fiducial gas model in Fig.~\ref{fig:overall} and two possible Milky Way analogs. Top row: \textit{Spitzer} IRAC 3.6 and 8$\;{\mu}m$ negative images of NGC~4565. Middle row: SDSS \textit{gri} images and \textit{Spitzer} IRAC 8$\;{\mu}m$ negative images of NGC~5746. These images are adapted from \citet{kor_ben_19}. Bottom row: WISE image of the Milky Way from \citet{nes_lan_16}, and our gas model viewed with an inclination angle of $87\degree$. Gas with density lower than $0.1\dunits$ is not shown. We stretch the images such that the b/p bulges and the inner rings have similar sizes for the three galaxies in the plot. 
\label{fig:mwanalog}}
\vspace{0.2cm}
\end{figure*}

\subsection{Bar-spiral interface and Molecular Ring}
\label{sec:barend}

\citet{reid_etal_19} found that their observed HMSFRs have significant non-circular motions ($\sim50\kms$) in a portion of the Perseus arm and near the bar ends. These large peculiar motions are believed to be related to the dynamics of the Galactic bar and spiral arms. \citet{baba_etal_18} proposed that a disrupting spiral arm is able to explain the peculiar motions in the Perseus arm, while no previous models can qualitatively reproduce the large peculiar motions at the bar ends.

We show in Fig.~\ref{fig:bessel} that our model can explain the observed peculiar motions at the bar end by introducing a bar-spiral interface term described by Eq.~\ref{eq:bs}. The left panel of Fig.~\ref{fig:bessel} plots the gas surface density of our fiducial model with $\Omb=37.5\freq$; the overlaid grey points are the observed HMSFRs described in \S\ref{sec:maserconstraints}. Eq.~\ref{eq:bs} is mainly responsible for generating the nearly vertical gas spur (shock) at $x\sim-2\kpc$ and $y\sim-4\kpc$ to $-6\kpc$ (and its counterpart at the other side of the Galaxy). The observed HMSFRs seem to be preferentially located around this spur, implying that it is a possible birth place. The peculiar motions of the observation and the model are shown in the middle and right panels of Fig.~\ref{fig:bessel}, respectively. Since our model does not include mechanisms to form HMSFRs, we assume the underlying gas flows have similar kinematics to the HMSFRs and thus plot the peculiar motion of the model gas at the location of the observed HMSFRs. The peculiar motions are calculated by subtracting the circular rotation velocity (Fig.~\ref{fig:rotcurve}) from the 3D velocity for both our model and observations. The color of the arrows in the right panel quantifies the velocity difference $v_{\rm diff}$ between the model and observations for each HMSFRs, which is defined as:
\begin{equation}
\label{eq:vdiff}
v_{\rm diff} = \sqrt{(\tilde{v}_{\rm R, obs.}-\tilde{v}_{\rm R, model})^2+(\tilde{v}_{\rm \varphi, obs.}-\tilde{v}_{\rm \varphi, model})^2}, 
\end{equation}
where $\tilde{v}_{\rm R}$ and $\tilde{v}_{\rm \varphi}$ are the peculiar motions along $R$ and $\varphi$ directions. Although there are still discrepancies for a few individual points, the overall patterns of the the model and data are similar, with a typical $v_{\rm diff}$ around $20\kms$.

We focus on the 26 HMSFRs inside the blue dashed box in the middle and right panels where the local potential is more affected by Eq.~\ref{eq:bs}. The HMSFRs in this region have a mean $\tilde{v}_{\rm R}$ of $-21.5\kms$ and a mean $\tilde{v}_{\rm \varphi}$ of $8.6\kms$, while for the same locations our gas model gives $-12.0\kms$ and $7.5\kms$, respectively. In general, gas is moving inwards and is faster than the local circular motion in this region. The reason for such a peculiar motion pattern is due to the bar-spiral interface that creates a local potential minimum around the bar ends. Gas is first accelerated when entering the potential minimum, but as the bar-spiral interface co-rotates with the bar, gas is then trapped around this region and turns into a local shock feature that is nearly radial. Since the post-shock region should be subsonic (in our case $\le10\kms$), the flow becomes mostly radial after passing the shock. The gas is then compressed at the shock front and is prone to the formation of HMSFRs. The model without the bar-spiral interface cannot form such a pattern (see Fig.~\ref{fig:bseffect}). Our results demonstrate that the potential related to the bar-spiral interface may need to be taken into account to study the kinematics of gas and stars in this region. We note the shape and amplitude of the spiral arms are still uncertain around the bar end. In principle, a strong arm with a large pitch angle in this region may also cause similar motions as the bar-spiral interface does, but this may imply a different spiral pattern compared to BeSSeL observations.

\citet{hilmi_etal_20} also studied the bar-spiral interaction and proposed that the length of the Milky Way bar may be over-estimated by $1-1.5\kpc$, and its pattern speed may be under-estimated by $5-10\freq$ compared to the time-averaged value. The reason they argued is because the Scutum-Centaurus-OSC\footnote{OSC stands for Outer-Scutum-Centaurus, see the definitions in \citet{reid_etal_19}.} arm may be connected to the near half of the bar at present. We find in our configuration that the bar-spiral interface indeed connects the bar and the Scutum arm, but it dominates only the local gas kinematics (i.e. around the bar ends). Other gas features like the 3-kpc arms are largely unaffected by the presence of this feature (e.g. Fig.~\ref{fig:3kpccomp}). It is possible that the bar-spiral interface needs to be considerably more massive than our case to alter the bar properties. However, the gas itself in this region may be sufficient to generate a local potential minimum that could explain the observed peculiar motions. The gas mass inside the purple box shown in the left panel of Fig.~\ref{fig:bessel} is $1.94\times10^7\Msun$, which is very close to the mass of the bar-spiral interface ($2.21\times10^7\Msun$) introduced by Eq.~\ref{eq:bs}. Further work is still needed to investigate in more detail the mass distribution around the bar end.   

%fig10 
\begin{figure*}[!t]
\includegraphics[width=1.0\textwidth]{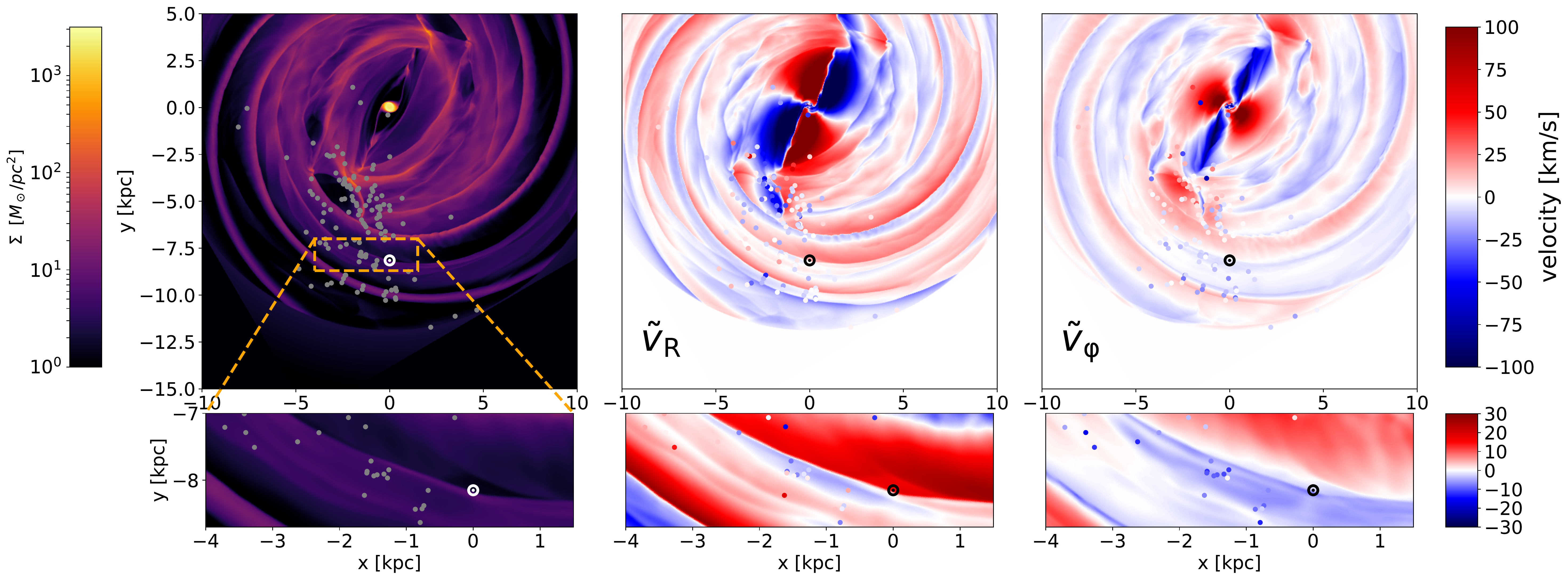}
\caption{Kinematic maps of the fiducial model with $\Omb=37.5\freq$. Left panels: gas surface density overlaid with the locations of HMSFRs (grey dots). The Sun is shown by the solar symbol. Middle panels: Galactocentric residual radial velocity $\tilde{V}_{R}$ in the model averaged in the $z$ direction. The dots represent the locations of HMSFRs with colors indicating the observed peculiar velocities in $R$ direction. Right panels: similar as the middle panels, but for Galactocentric residual azimuthal velocity $\tilde{V}_{\rm \varphi}$. The residual velocity maps are obtained by subtracting the circular rotation velocity from the gas velocity field at $t=440\Myr$. The bottom panels are the zoom-in views around the SNd defined by the orange dashed box in the upper left panel. 
\label{fig:vrvtsun}}
\vspace{0.2cm}
\end{figure*}

One may wonder to what extent HMSFRs trace the underlying gas flows. This may depend on their lifetime and the ratio of mass density in HMSFRs to the surrounding gas, although these properties may not be easily measured. \citet{reid_etal_19} found that the observed HMSFRs are relatively concentrated near the high-density gas ridges in the \lvplot, with a typical dispersion of $10-20\kms$. Assuming the tangential velocity has a similar scatter, this leads to a total velocity offset of $15-30\kms$. The simulations by \citet{baba_etal_09} suggested a comparable velocity difference between the motions of dense gas and young stars. The overall 135 HMSFRs have an average $v_{\rm diff}$ of $24.5\kms$ compared to our gas model shown in Fig.~\ref{fig:Omb37}, roughly consistent with the above differences. We therefore conclude that HMSFRs probably trace the underlying global gas flows reasonably well. A careful work on the formation of HMSFRs in disk galaxies is required to understand this problem further. 

Another interesting feature in this region is the elliptical gas ring surrounding the bar. This ring may contribute to part of the Molecular Ring \citep{dame_etal_01,romand_etal_10} which is the high-density and low $v_{\rm LOS}$ strip between $\abs{l}\la30\degree$, and the 3-kpc arms in the \lvplot\ (see the middle panels of Fig.~\ref{fig:surflv}). One question raised by \citet{kor_ben_19} is whether our Milky Way is an SB(r)bc galaxy with a gaseous ``inner ring''. The answer is probably yes according to our gas models. Note that the inner ring in our model is composed not only of the 3-kpc arms \citep[e.g. suggested by][]{sevens_etal_01}, but also part of the Norma arm and the bar-spiral interfaces that connect to the outer spirals. We show in Fig.~\ref{fig:mwanalog} that when inclined at an angle of $87\degree$, the inner ring of our fiducial gas model looks quite similar to those observed Milky Way analogs, such as NGC 4565 and NGC 5746 \citep{kor_ben_19}.

\subsection{The Solar Neighbourhood (SNd) and the Local Arm}
\label{sec:locarlarm}

The Local arm was previously assumed to be a spur instead of a major arm as the latter explanation is not favored by the density wave theory \citep{yuan_69}. However, recent observations seem to suggest the Local arm has a spatial extent of at least $\sim6\kpc$, and its pitch angle, width, and star formation rate are comparable to those of the major arms \citep[e.g.][]{xu_etal_13,xu_etal_16}. If the Local arm is not a spur, then a theoretical explanation is needed to produce this feature together with the other four major arms. 

We present one possible scenario to form the Local arm. In our gas models the Local arm is induced by the spiral arm potential described by Eq.~\ref{eq:spiral}. Note that we only include a 4-arm spiral perturbation potential, but the corresponding gas flows form a 6-arm pattern spontaneously. The development of higher-order spiral features (sometimes dubbed as ``branches'') has also been found in previous studies \citep[e.g.][]{martos_eta_04,pettit_etal_14,few_etal_16}. We believe the Local arm in our models is related with $\Phi_6$, which has a clear variation around the solar radius as can be seen in Fig.~\ref{fig:frfphi}. The inner 4:1 resonance of the spiral is $6.8\kpc$ (Fig.~\ref{fig:rotcurve}), which may also help to shape the Local arm. We have verified that the Local arm co-rotate with the spiral potential ($\Omega_{\rm sp}=23\freq$) in our models, thus the bar is probably not relevant to the formation of the Local arm. The scenario present here is similar to that in \citet{pettit_etal_14} where the authors argued the Local and Outer arms may be branches generated by the spiral potential.

We show in Fig.~\ref{fig:vrvtsun} that the Local arm has a clear kinematic pattern in the SNd suggested by our models, despite that the local underlying force distribution in Fig.~\ref{fig:frfphi} is rather smooth. The left panels of Fig.~\ref{fig:vrvtsun} plot the gas surface density overlaid with observed HMSFRs (grey points), similar to Fig.~\ref{fig:bessel}. We further show the zoom-in plots around the SNd in the bottom panels. The Sun is located just inside the Local arm which has a pitch angle of $\sim12\degree$, similar to the value constrained by HMSFRs \citep{reid_etal_19}. In the middle and right panels we show maps of the residual velocity along the radial and azimuthal directions (i.e. $\tilde{V}_R$ and $\tilde{V}_{\rm \varphi}$), which are obtained by subtracting the circular rotation velocity (Fig.~\ref{fig:rotcurve}) from the gas velocity fields. The central quadrupole feature in the upper middle panel is evidently due to the bar, and each spiral has its kinematic signature (e.g. $\tilde{V}$ switches sign across spirals) in the residual velocity maps. Due to the presence of the Local arm, the gas in our model is moving outwards and is faster than the circular speed just inside the solar radius, with a typical amplitude of $10-20\kms$. This value is comparable to the gas velocities perturbed by the major spiral arms which are explicitly included in the underlying potential. The observed HMSFRs around the SNd (colored dots) are located mainly outside the solar circle, and they have negative $\tilde{V}_{\rm \varphi}\sim-10\kms$ \citep{xu_etal_13}. This seems to agree with the model prediction, but the $\tilde{V}_R$ pattern of the HMSFRs is not quite clear, probably due to the limited number of sources inside the solar circle where the outflow signal is strong. 

Interestingly, a stellar $\tilde{V}_R$ pattern similar to Fig.~\ref{fig:vrvtsun} has been reported by \citet{eilers_etal_20}, and they interpreted it as a dynamical effect of the stellar Local arm. The major differences between \citet{eilers_etal_20} and this work is we do not impose a local spiral potential, but the gaseous Local arm and its related kinematics in our model is spontaneously induced by the 4-armed pattern of the major stellar spiral arms. 

It is still unclear whether the Local arm has an old stellar counterpart. Recent work by \citet{miyach_etal_19} found a marginal overdensity of stars near the HMSFR-defined Local arm, with a slightly larger pitch angle. The locations of the young stars in Gaia eDR3 also display an arm feature near the Sun \citep{poggio_etal_21,xu_etal_21}. It is possible that these young stars are formed from the gas arm. Future Galactic surveys may provide a definitive answer on whether there is a massive stellar Local arm composed of old stars near the SNd or not.

\section{Discussion}
\label{sec:discussion}

\subsection{Moving Groups in the Solar Neighbourhood}

The gas kinematics prefer $\Omb=37.5-40\freq$ as suggested by our models. It is therefore straightforward to compare this value with other independent measurements. The moving groups in the SNd, especially the Hercules stream, have been extensively used to constrain the bar pattern speed for decades. \citet{dehnen_00} and \citet{antoja_etal_14} suggested the Hercules stream is an Outer Lindblad Resonance (OLR) signature of the bar if $\Omb$ is larger than $\sim50\freq$. The stream could also be reproduced with a bar pattern speed of $\sim40\freq$ by the orbits near the $\RCR$ of the bar \citep[e.g.][]{perez_etal_17,monari_etal_19,binney_20a,chi_sch_21}, and/or the 4:1 resonance of the bar \citep{hun_bov_18}. We have verified that our galactic potential in the fiducial models with $\Omb=37.5-40\freq$ also favors a CR-origin explanation, suggesting $\Omb=1.24-1.32~\Omega_0$ ($\Omega_0\equiv\ v_{0}/R_0$), consistent with $\Omb=1.27~\Omega_0$ obtained by \citet{trick_etal_21}. Note that the Hercules stream may be a bimodal or even trimodal structure as revealed in Gaia DR2 results \citep{gaiacoll_etal_18,trick_etal_19,asano_etal_20}. Only one of the branches shows clear phase space snail structure in the $z-V_{z}$ plane \citep{li_shen_20}. The model that includes the bar resonances alone is probably not enough to split the Hercules stream into multiple components with different vertical kinematics, hinting for a more complicated formation mechanism.

%fig10 
\begin{figure}[!t]
\includegraphics[width=0.5\textwidth]{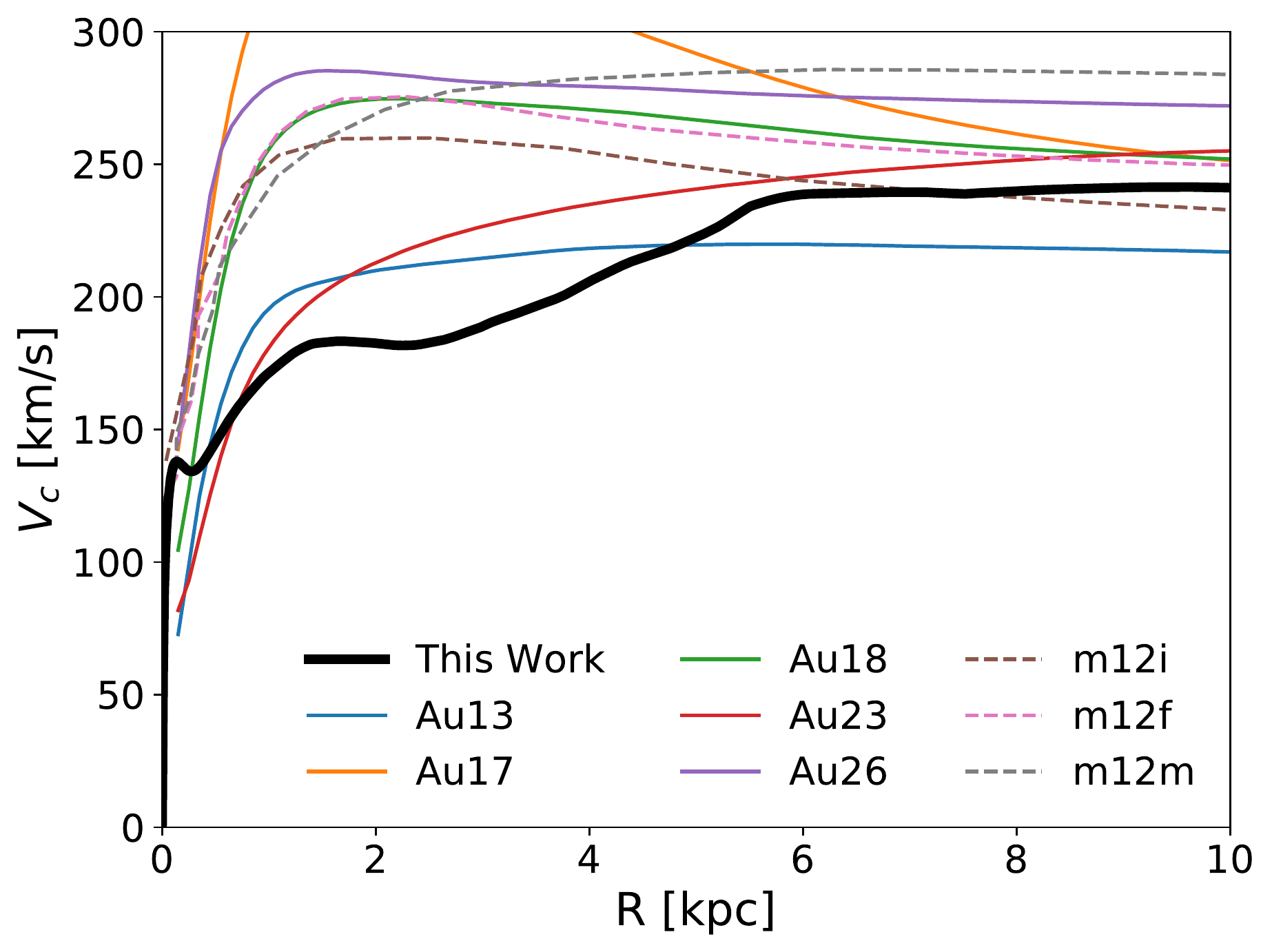}
\caption{Comparison of the rotation curve between our work (thick black line, same as the solid blue line in Fig.~\ref{fig:rotcurve}) and those from the Auriga (solid color lines) and LATTE simulations (dashed color lines).  
\label{fig:aurigacomp}}
\vspace{0.2cm}
\end{figure}

\subsection{Comparing the MW Rotation Curve with Cosmological Simulations}

Cosmological simulations offer important insights for understanding the formation and mass assembly history of the Milky Way. For example, the Auriga project \citep{grand_etal_17} and the LATTE suite \citep{wetzel_etal_16} have provided a number of simulated Milky Way analogs with diverse formation histories in a fully cosmological context, and the models have been widely used in various aspects of Milky Way studies \citep[e.g.][]{fragko_etal_20,cautun_etal_20,sander_etal_20,grand_etal_21}.

Here we constrained the MW potential and rotation curve by modelling different observational data, and it is interesting to compare it with the rotation curves of these cosmological MW analog models. The dynamical bar models of P17 with which we started our analysis used star counts and stellar kinematics in the bulge for an accurate mass determination in the central $2\kpc$, and inferred the rotation curve near the Sun ($R=6-8\kpc$) from terminal velocities assuming circular orbits. Inside $6\kpc$ the influence of the Galactic bar cannot be neglected, and therefore their rotation curve between both regions was based on fitting a dark matter mass model, i.e. had a model-dependent component. In this paper we showed that with relatively small modifications to the P17 rotation curve we can reproduce the gas flow and in particular the terminal velocities from the solar radius all the way into the bulge region. The new rotation curve is therefore based on observational data at all these radii, and is therefore more secure than that of P17. We reiterate here that the observed high gas LOS velocities ($\sim250\kms$ within $\abs{l}\la10\degree$) are mostly caused by the non-circular flows along the bar, and do not represent the underlying mass distribution in the central region \citep[e.g., see][]{binney_etal_91,chemin_etal_15}.

In Fig.~\ref{fig:aurigacomp} we compare the rotation curve (i.e. the mass distribution) inferred by the current work (black line) with 5 Auriga models from \citet{fragko_etal_20} (colored solid lines) and 3 LATTE models from \citet{sander_etal_20} (colored dashed lines). The 5 Auriga models are strongly barred galaxies with b/p bulges while the bars in the 3 LATTE galaxies are weaker and younger \citep{debatt_etal_19}. Although the two suite zoom-in simulations are different in many aspects (e.g. initial conditions, the merger/mass assembly history, detailed modelling of sub-grid physics, numerical resolution, etc.), the rotation curves of the MW analogs in these simulations are all relatively high ($\geq200\kms$) in the central $R\la2\kpc$, due to the contribution of a compact bulge. This peak is clearer with higher numerical resolution, probably caused by the non-linear star formation history which is sensitive to the resolution and the sub-grid physics modules implemented \citep[see the discussion in ][]{grand_etal_17}.

Despite these uncertainties, the combined stellar and gas dynamical models for the MW suggest that in the inner $2\kpc$ of the Milky Way the mass distribution is less concentrated. The dynamical mass in the bulge volume (i.e. $\lesssim2\kpc$) estimated by P17 is $1.85\pm0.05\times10^{10}\Msun$, and this already results in a more gently rising rotation curve reaching $\sim180\kms$ at $R\sim2\kpc$. But also at larger radii (e.g. $\sim5\kpc$) the inferred rise in the MW rotation curve is slower. Such a rotation curve seems less common in the cosmological simulations, and an interesting question is therefore under what conditions disk galaxies are formed with a slowly rising rotation curve (or a less compact central region). The answer would provide us more clues to a better understanding of the MW's formation history. 

\subsection{Future improvements}
\label{sec:futurework}

Our current gas models carefully compute the flow of an ideal, isothermal gas, in set of fixed MW potential models. However, we neglect more complicated physical processes like radiative cooling, star formation, and stellar feedback that may affect many properties of the gas flows. We restrict ourselves to isothermal gas because we would like to understand first how gas with a typical velocity dispersion of $\sim10\kms$ evolves in a MW-like potential, which highlights the effects of gravity instead of local gas physics. A clear improvement is therefore to incorporate additional physics in the current fiducial models, similar to other recent attempts to study the gas structures in the MW \citep[e.g.][]{bab_kaw_20,pettitt_etal_20,reissl_etal_20}.

Another improvements is to use a more accurate potential. We see in \S\ref{sec:ombdetermin} that the P17 potentials may underestimate the stellar disk and/or the dark matter contributions inside $R\sim5\kpc$, which motivates us to include $\Phi_{\rm pl}$ to better match the terminal velocities. The mass distribution outside the solar circle is not well-constrained by the current gas models either. Besides, our models assume a fixed potential that neglects the possible perturbations of satellite galaxies \citep[e.g.][]{antoja_etal_18,blandh_etal_19,li_shen_20}. These effects on the gas dynamics in the MW may need to be better investigated in future studies.

\section{Summary}
\label{sec:summary}

We use gas dynamical models to study the gravitational potential and the bar pattern speed of the Milky Way. The basis Galactic potentials are from the stellar dynamical models in P17, which are well-constrained by the star counts and stellar kinematics. Our gas models provide independent and additional constraints compared to the stellar models, and further improve our understandings on the Galaxy. The main findings in this work are summarized as follows:

(1) Our model favors a bar pattern speed in the range of $37.5-40\freq$ ($\RCR=6.0-6.4\kpc$) based on the diagnostics of the observed 3-kpc arms and the forbidden velocity region.

(2) Our barred potential with a nuclear stellar disk of $6.9\times10^8\Msun$ from Jeans modelling results can generate a central gas disk with a similar size and kinematics compared to the observed CMZ.

(3) A localized bar-spiral interface with a mass of $0.44\times10^8\Msun$ may be helpful to reproduce the observed large peculiar motions of HMSFRs around the bar end. It also helps to create a gaseous ``inner ring'' that is similar to external MW analogs.

(4) The Local arm may be a "branch" that is induced by the 4-arm spiral potential, as has been suggested by previous studies. In addition, we show this can result in clear gas kinematic patterns around the SNd which agree well with observational data.

(5) Our fiducial models can generate a steady gas flow pattern that reproduce most of the observed $(l,v)$ features, the terminal velocities, and the peculiar motions of HMSFRs. The rotation curve of the fiducial models has a gently rising shape within $R\sim5\kpc$ instead of a clear peak feature in the central region. The observed high gas LOS velocities ($\sim250\kms$ within $\abs{l}\la10\degree$) are mostly caused by the non-circular flows along the bar, which cannot be used to trace the real mass distribution.

\software{
{\tt Athena++} \citep{stone_etal_20},
Astropy \citep{astropy:2013,astropy:2018},
NumPy \citep{2020NumPy-Array},
SciPy \citep{2020SciPy-NMeth},
Matplotlib \citep{4160265},
Jupyter Notebook \citep{Kluyver2016jupyter}
similaritymeasures \citep{jekel_etal_19}
}

\begin{acknowledgments}
We thank the anonymous referee for suggestions that help to improve the presentation of the paper. ZL would like to thank Robert Grand for sharing the Auriga rotation curves, and Jonathan Henshaw for sharing the NH$_3$ data, and Xiangcheng Ma for helpful discussions. 
The research presented here is partially supported by the National Key R\&D Program of China under grant No. 2018YFA0404501; by the National Natural Science Foundation of China under grant Nos. 12025302, 11773052, 11761131016; by the ``111'' Project of the Ministry of Education of China under grant No. B20019; and by the Chinese Space Station Telescope project. JS acknowledges the support of a \textit{Newton Advanced Fellowship} awarded by the Royal Society and the Newton Fund. OG acknowledges the support by Deutsche Forschungsgemeinschaft under grant GZ GE 567/5-1. This work made use of the Gravity Supercomputer at the Department of Astronomy, Shanghai Jiao Tong University, and the facilities of the Center for High Performance Computing at Shanghai Astronomical Observatory.
\end{acknowledgments}

\appendix

\restartappendixnumbering

\section{Different models}

\begin{figure*}[!t]
\includegraphics[width=1.0\textwidth]{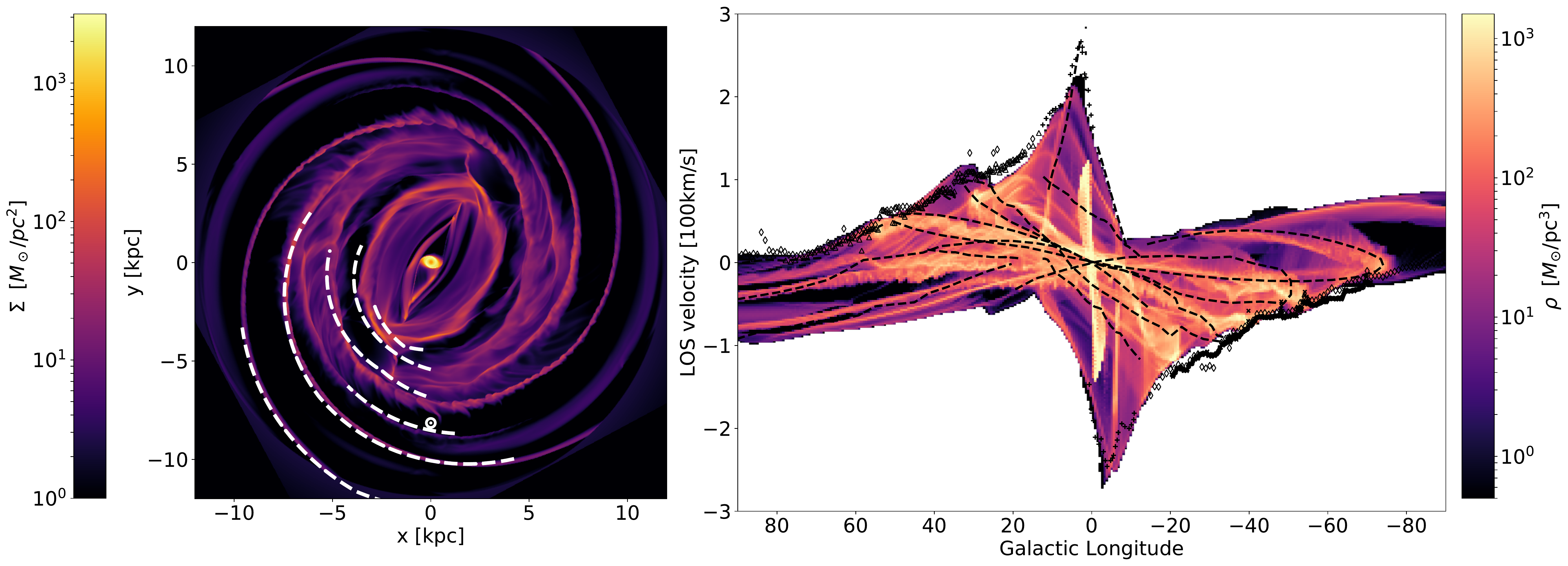}
\caption{Gas model using the same potential and bar pattern speed as Fig.~\ref{fig:Omb37}, but with an effective sound speed of $5\kms$. 
\label{fig:cs5}}
\vspace{0.2cm}
\end{figure*}

\begin{figure*}[!t]
\includegraphics[width=1.0\textwidth]{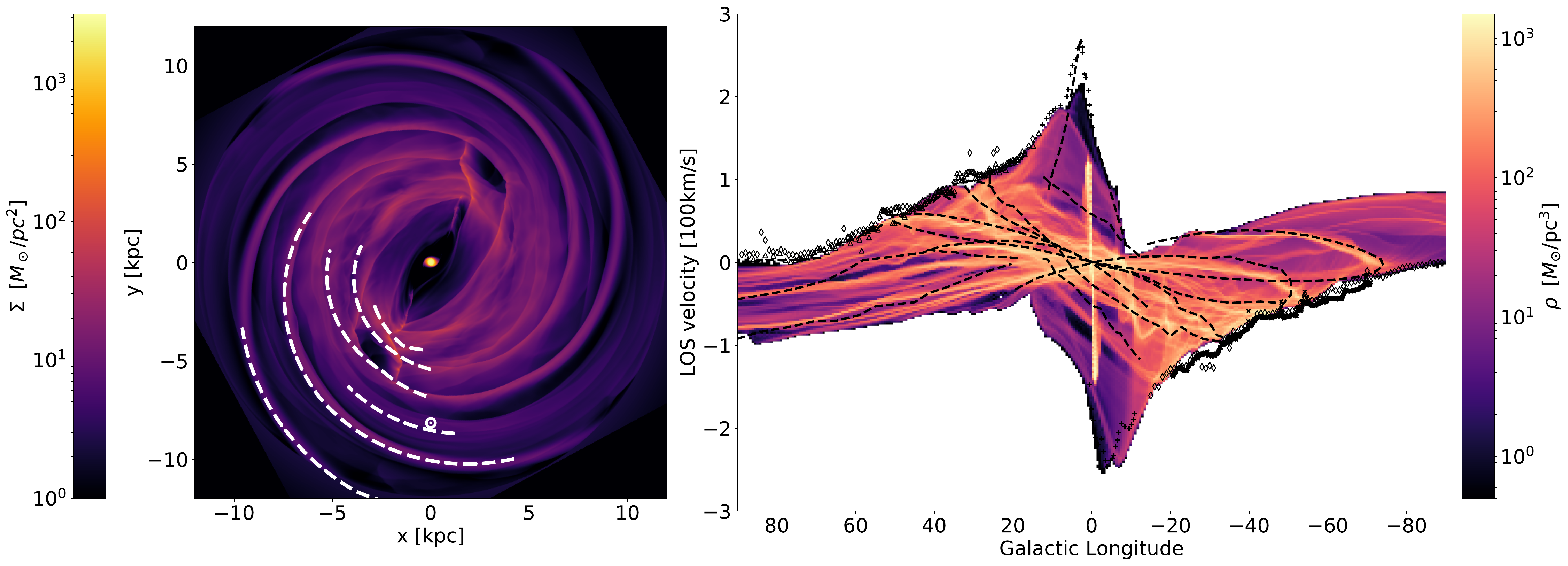}
\caption{Gas model using the same potential and bar pattern speed as Fig.~\ref{fig:Omb37}, but with an effective sound speed of $15\kms$.  
\label{fig:cs15}}
\vspace{0.2cm}
\end{figure*}

We present the results of different gas models in the appendix. Fig.~\ref{fig:bseffect} shows the effects of different bar-spiral interface. Figs.~\ref{fig:cs5} and \ref{fig:cs15} shows the effects of different gas effective sound speed.

%figA1 
\begin{figure*}[!t]
\includegraphics[width=1.0\textwidth]{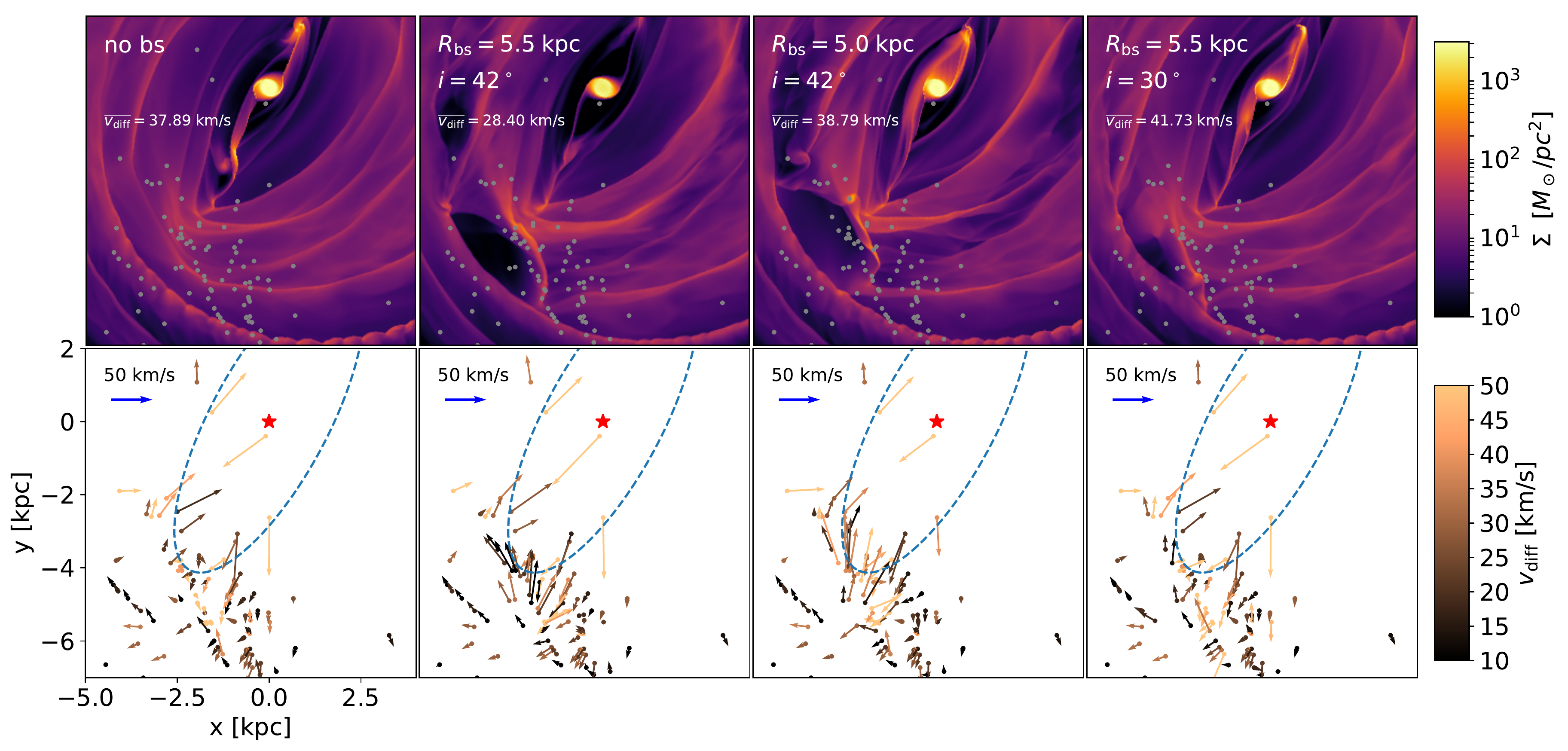}
\caption{The effects of bar-spiral interface with different parameters. From left to right: the model without a bar-spiral interface; the model with the bar-spiral interface described in \S\ref{sec:bspotential} (i.e. the fiducial model); the model with a bar-spiral interface but adopting a different location of $R_{\rm bs}=5.0\kpc$; the model with a bar-spiral interface but adopting a different pitch angle of $i=30\degree$. Lines and points are the same as in Fig.~\ref{fig:bessel}. $\overline{v_{\rm diff}}$ is the average of $v_{\rm diff}$ for the 26 masers within the white dashed box in Fig.~\ref{fig:bessel}.
\label{fig:bseffect}}
\vspace{0.2cm}
\end{figure*}

%figA2 

%figA3 

\bibliographystyle{aasjournal}
\bibliography{gasdynamics}

\end{document}